\documentclass[preprint]{ptephy_v1}

\preprintnumber{XXXX-XXXX} 

\usepackage{amsmath}
\usepackage{comment}
\usepackage[separate-uncertainty=true]{siunitx}
\usepackage{lastpage}
\usepackage{booktabs}
\usepackage{multirow}
\usepackage{lineno}
\usepackage{color}
\usepackage{todonotes}
\usepackage{algorithm}
\usepackage{algpseudocode}
\usepackage{aas_macros} 
\usepackage{hyperref}
\hypersetup{
colorlinks=true,
urlcolor=cyan,
citecolor=blue,
linkcolor=red,
filecolor=magenta, 
pdftitle={Extended Delta-map},
pdfauthor={Yuto Minami}
}

\newcommand*{\bm}[1]{\mathbf{#1}}%
\newcommand{\T}{\mathsf{T}}
\newcommand{\Dfg}{\tilde{\bm{D}}}
\newcommand{\Dcmb}{ {\mathbf{D}^\mathrm{CMB} } }
\newcommand{\Sfg}{{\mathbf{S}^{f}}}
\newcommand{\Scmb}{\mathbf{S}^\mathrm{CMB}}
\newcommand{\Socmb}{ {\mathbf{S}_0^\mathrm{CMB} } }
\newcommand{\sfg}{ {\vec{\tilde{s} }_f} }
\newcommand{\Npix}{ {N_\mathrm{pix}} }

\begin{document}

\title{Extended Delta-map: a map-based foreground removal method for CMB polarization observations}


\author[1]{Yuto Minami}
\affil[1]{Research Center for Nuclear Physics, Osaka University, Ibaraki, Osaka 567-0047, Japan \email{yminami@rcnp.osaka-u.ac.jp} }

\author[2,3,4]{Kiyotomo Ichiki}
\affil[2]{Kobayashi-Maskawa Institute for the Origin of Particles and the Universe, Nagoya University, Furo-cho, Chikusa-ku, Nagoya, Aichi 464-8602, Japan}
\affil[3]{Department of Physics and Astrophysics, Nagoya University, Furo-cho, Chikusa-ku, Nagoya, Aichi 464-8602, Japan \email{ichiki.kiyotomo.a9@f.mail.nagoya-u.ac.jp}}
\affil[4]{Institute for Advanced Research, Nagoya University, Furo-cho, Chikusa-ku, Nagoya, Aichi 464-8602, Japan \email{ichiki.kiyotomo.a9@f.mail.nagoya-u.ac.jp}}


\begin{abstract}%
In order to extract information about inflationary gravitational waves using $B$-mode patterns of cosmic microwave polarization anisotropy,
we need to remove the foreground radiation from the Milky Way.
In our previous delta-map method for foreground removal,
the number of observation bands was limited to the number of parameters of the assumed foreground model, and therefore it was difficult to improve the sensitivity by increasing the number of observation bands.
Here, we extend the previous method so that it can be adapted to an arbitrary number of observation bands.
Using parametric likelihood and realistic foreground and CMB simulations,
we show that our method can increase the sensitivity to the tensor-to-scalar ratio $r$ without inducing any significant bias.
\end{abstract}

\subjectindex{E63, F11, F14}

\maketitle

\section{Introduction\label{sec:intro}}
Precise measurements of linear polarization patters of the cosmic microwave background (CMB) provide us with plentiful information of our Universe~\cite{Komatsu:2014ioa,Aghanim:2018eyx, POLARBEAR:2022dxa, Polarbear:2020lii, ACT:2020gnv, SPT:2019nip,Dutcher:2021vtw,BICEP:2021xfz,SPIDER:2021ncy}.
In particular, a measurement of $B$-mode pattern in the CMB polarization anisotropies is a key to detecting primordial gravitational waves (PGWs) expected in the inflation scenario of the early Universe~\cite{Grishchuk:1974ny,Starobinsky:1979ty,Seljak:1996gy,Kamionkowski:1996zd, Kamionkowski:2015yta}. 
However, celestial sources also emit microwave photons with linear polarization that mimic the cosmological $B$-mode, the so-called foreground emission, which disturbs the precise measurement of the CMB $B$-mode~\cite{Ichiki:2014qga}.
The two representative foreground emissions in CMB polarization measurements are synchrotron and thermal dust emissions, which have different spectral energy distribution from the black body and dominate at lower and higher frequencies, respectively.
Therefore, we can remove the foreground emissions using multiple-band observations owing to their frequency dependences.

Many foreground removal and component separation methods exist to extract the cosmological $B$-mode signal, e.g.,
Commander~\cite{Eriksen:2007mx}, SEVEM \cite{2012MNRAS.420.2162F}, SMICA~\cite{Delabrouille:2002kz}, NILC~\cite{2009A&A...493..835D}, and GNILC~\cite{2011MNRAS.418..467R}.
Among them, we proposed a method called ``delta-map'' based on a linear combination of the observed maps to remove the foreground components~\cite{Ichiki:2018glx}.
This method allows the directional dependence of the frequency spectrum of foreground emissions up to the first order to be taken into account by considering the differences in the observation maps
\footnote{A similar perturbative approach in multipole space has also been investigated in Refs.~\cite{Chluba:2017rtj,Mangilli:2019opl,Vacher:2021heq,Vacher:2022xdw}.}.
Because the method uses one additional frequency map to eliminate the directional dependence of one foreground parameter,
the usable number of
maps
is limited by the number of foreground parameters in the assumed model. 
Specifically, suppose that we assume one foreground model with $N$ parameters and CMB signal in all the multi-frequency maps.
In this case, the delta-map method requires one map for the CMB, one map for the zeroth (or spatially uniform) foreground parameter, and $N$ maps for the first order expansion of the $N$ parameters;
we need exactly $( 2 + N )$ maps.
This prevents us from obtaining improvement of the sensitivity in observation by increasing the number of observing bands.
More concretely, if we consider a power-law synchrotron foreground model with one parameter $\beta_s$ and a one-component dust model with two parameters $\beta_d$ and $T_d$,
the number of observation bands should be exactly six.
This is a waste of resources since some future CMB missions plan to have plenty of bands, e.g., LiteBIRD will have $15$ bands \cite{Hazumi2019,LiteBIRD:2022cnt}.
In this paper, we improve the method by constructing a parametric likelihood in a Bayesian way so that more observation bands can be used.

The rest of this paper is organized as follows.
In Sect.~\ref{sec:method}, we first review the previous delta-map method that is based on the linear combination,
and then derive a new delta-map method based on a parametric likelihood.
In Sect.~\ref{sec:sim}, we explain the foreground models and simulations used in this work.
We show the results of the measurements of CMB parameters with some sky simulation setups in Sect.~\ref{sec:results}.
In Sect.~\ref{sec:conclusion}, we discuss the results and conclude this work.

\section{Methodology\label{sec:method}}
We first introduce the delta-map method described in Ref.~\cite{Ichiki:2018glx}.
We decompose a linear polarization signal of a microwave component at a frequency, $\nu$,
from a line of sight, $\hat{n}$,
to two orthogonal Stokes parameters, $Q(\hat{n})$ and $U(\hat{n})$,
which we observe in thermodynamic temperature units.
Let us vectorize all $Q (\hat{n})$ and $U(\hat{n})$ from all $\hat{n}$ in use and concatenate them into one column vector as
\begin{align}
	\vec{s} = 
	\left( Q(\hat{n}_1 ), \ldots, Q(\hat{n}_{\Npix} ) , U(\hat{n}_1), \ldots,  U(\hat{n}_{\Npix } )    \right)^\T,
\end{align}
where $\Npix$ is the number of pixels in use and the subscript $\T$ represents transpose.

The CMB signal is common in any frequency band except for beam effects of instruments, and can be written as $\vec{s}_\mathrm{CMB} ( \nu ) = \vec{s}_\mathrm{CMB} $.
We assume that each frequency has each independent Gaussian noise, $\vec{s}_{N}$.
A foreground signal varies in frequency space according to $N$ parameters that vary over the sky,  $\vec{p}^{I}\, (I = 1,2, \cdots, N)$, as 
\begin{align}\label{eq:sig_freq}
	\vec{s}_f (\nu) = \textsl{g}_\nu \bm{D}_\nu ( \vec{p}^{I}) \vec{s}_b ,
\end{align}
 where 
$\vec{s}_b$
 is a signal vector at a pivot frequency in brightness temperature units, $\bm{D}_\nu( \vec{p}^{I})$ is a diagonal matrix with the same dimension of $\vec{s}_f$, and  $\textsl{g}_\nu$ is the conversion factor from the brightness temperature to the CMB thermodynamic temperature given by 
\begin{align}
 	\textsl{g}_\nu \equiv \frac{(e^x -1 )^2}{ e^x x^2} \text{ with } x\equiv  \frac{h\nu}{k_B T_\text{CMB}}~.
\end{align}

In this paper, we use power-law synchrotron and one-component modified black body (1MBB) models to fit synchrotron and dust foreground emissions, respectively.
The functional form of one pixel of $\bm{D}_\nu ( \vec{p}^{I})$ for power-law synchrotron is
\begin{align}
    D_\nu^s(\beta_s (\hat{n})) = \left(\frac{\nu}{\nu_{s_*}}\right)^{\beta_s(\hat{n}) },
\end{align}
where $\nu_*$ is the reference frequency, which we set as $\nu_{s_*}=\SI{23}{GHz}$, and $\beta_s (\hat{n})$ is the synchrotron spectral index.
The one for 1MBB is 
\begin{align}
    D_\nu^d(T_d (\hat{n}), \beta_d(\hat{n}) ) = \left(\frac{\nu}{\nu_{d_*}}\right)^{\beta_d(\hat{n}) + 1} \frac{e^{x_{d*}(\hat{n})} - 1 }{e^{x_{d}(\hat{n})} - 1},
\end{align}
where $x_{d*} (\hat{n})  \equiv  \frac{h\nu_{d_*} }{k_B T_d(\hat{n})}$,
$x_d (\hat{n})  \equiv  \frac{h\nu}{k_B T_d(\hat{n})}$,
$T_d (\hat{n})$ is the dust temperature,
$\beta_d (\hat{n})$ is the dust spectral index,
and we set $\nu_{d_{*}} = \SI{353}{GHz} $.

The basic idea of the delta-map method is to consider a spatial variation of foreground signal parameters up to the first-order expansion\footnote{The standard deviation of directional variation of foreground parameters considered in this paper is less than $10\%$ of the spatial uniform parameter. We can assume second-order variation to be less than $1\%$; hence we neglect higher-order terms.} as
\begin{align}\label{eq:1stexpansion}
	\textsl{g}_\nu\bm{D}_\nu ( \vec{p}^{I}) \vec{s}_b &=  \textsl{g}_\nu D_\nu (\bar{p}^I) \bm{I}  \vec{s}_b + \textsl{g}_\nu\textstyle{\sum}_{I=1}^{N} D_\nu,_{p^I} (\bar{p}^I) \bm{I} (\delta\vec{p}^{I}\circ  \vec{s}_b)+ \mathcal{O}({ \delta\vec{p}^{I} }^2)~, 
\end{align}
where $\bar{p}^I$ represents the mean value of the parameter $\vec{p}^I$ over the sky, $\bm{I}$ is an identity matrix and ``$\circ$'' means Hadamard product.
Hereafter, we consider two representative foreground components:
synchrotron radiation and thermal dust emission.
With these two foreground components, the foreground signal is written as
\begin{align}\begin{split}
	\textsl{g}_\nu\bm{D}_\nu ( \vec{p}^{I}) \vec{s}_f
	&=  
	\textsl{g}_\nu D_\nu^{d} (\bar{p}^I) \bm{I} \vec{s}_{d} + \textsl{g}_\nu\sum_{I=1}^{N_d} D_\nu^{d},_{p_d^I} (\bar{p}^I) \bm{I} (\delta\vec{p}_d^{I}\circ \vec{s}_{d})+ \mathcal{O}({ \delta\vec{p}_d^{I} }^2)\\
	&+ \textsl{g}_\nu D_\nu^{s} (\bar{p}^I) \bm{I} \vec{s}_{s} + \textsl{g}_\nu \sum_{I=1}^{N_s} D_\nu^{s},_{p_s^I} (\bar{p}_s^I) \bm{I} (\delta\vec{p}_s^{I}\circ \vec{s}_{s}) + \mathcal{O}({ \delta\vec{p}_s^{I} }^2)\\
	&=  \Dfg_\nu \sfg 
	,
\end{split}\end{align}
where the superscripts and subscripts ``$s$'' and ``$d$'' denote synchrotron and dust emission foreground components, respectively,
and $N_s$ and $N_d$ denote the numbers of parameters for synchrotron and dust foreground emissions, respectively.
Here $\Dfg_\nu$ represents frequency dependence of the signals up to the first-order $\sfg$ and given by
\begin{align}\label{eq:Dfg_nu}
\Dfg_\nu = \begin{pmatrix}
\textsl{g}_\nu D_\nu^d (\bar{p}_d^I) \bm{I} &  \textsl{g}_\nu D_\nu^d,_{p_d^1} (\bar{p}_d^1)  \bm{I}& \cdots & \textsl{g}_\nu D_\nu^d,_{p_d^{N_d} } (\bar{p}_d^{N_d})  \bm{I} &
  \textsl{g}_\nu D_\nu^s (\bar{p}_s^I) \bm{I} &  \textsl{g}_\nu D_\nu^s,_{p_s^1} (\bar{p}_s^1) \bm{I} & \cdots & \textsl{g}_\nu D_\nu^s,_{p_s^{N_s}} (\bar{p}_s^{N_s})  \bm{I}
\end{pmatrix},
\end{align}
and, at the last line, we define the foreground signal vector as 
\begin{align}
	\sfg \equiv \begin{pmatrix}
		\vec{s}_{d}\\ \delta\vec{p}_d^{1}\circ \vec{s}_{d} \\ \vdots\\ \delta\vec{p}_d^{N} \circ \vec{s}_{d} \\
		\vec{s}_{s} \\ \delta\vec{p}_s^{1}\circ \vec{s}_{s} \\ \vdots\\ \delta\vec{p}_s^{N} \circ \vec{s}_{s} 
	\end{pmatrix}.
\end{align}
In summary, we assume the observed data can be decomposed as
\begin{align}
	\vec{m}(\nu) = \vec{s}_\mathrm{CMB} + \Dfg_\nu \sfg + \vec{s}_{N}(\nu)~.
\end{align}

We need to observe the sky at multiple frequencies to remove foreground components using its frequency dependence.
When we observe the sky at $N_\nu$ frequencies, the total data can be expressed as
\begin{align}\label{eq:map_components}
	\vec{m} = \Dcmb \vec{s}_\mathrm{CMB} + \Dfg \sfg + 
\begin{pmatrix}
	\vec{s}_{N} (\nu_1)\\ \vdots \\ \vec{s}_{N} (\nu_{N_\nu})
\end{pmatrix},
\end{align}
where 
\begin{align}
	\vec{m} \equiv	\begin{pmatrix}
		\vec{m}(\nu_1 )\\ \vdots\\ \vec{m}(\nu_{N_\nu} ) 
	\end{pmatrix},
\end{align}
\begin{align}
	\Dcmb \equiv \begin{pmatrix}
		\bm{I}\\ \vdots \\ \bm{I}
	\end{pmatrix} \quad \text{ and }  \quad 	\Dfg \equiv \begin{pmatrix}
	\Dfg_{\nu_1} \\ \vdots \\ \Dfg_{\nu_{N_\nu}}
\end{pmatrix}.
\end{align}
This is the baseline expression of the ``delta-map'' method.

In Ref.~\cite{Ichiki:2018glx},
one frequency, $\nu_1$, was chosen as the CMB channel and other channels with weights $\vec{\alpha}^\T = (\alpha_{\nu_2}, \ldots, \alpha_{\nu_{N_\nu}})$ were added to remove foreground contributions and to have a cleaned CMB map as
\begin{align}\label{eq:cleanedCMBalpha}
	\vec{m}_\mathrm{CMB} = 
	\frac{\vec{m}_\mathrm{\nu_\mathrm{CMB}} +
		 \sum_{i = \nu_2}^{\nu_{N_\nu } } \alpha_i \vec{m}_\mathrm{\nu_i}
	 }{1 + \sum_{i = \nu_2}^{\nu_{N_\nu}} \alpha_i  },
\end{align}
with 
\begin{equation}
[1, \vec{\alpha}^\T ] \tilde{\bm{D}}  = \bm{O}~.
\end{equation}
The above equation can be solved only when the number of frequencies, $N_\nu$,
is equal to the number of degrees of freedom, $(N_d+1) + (N_s+1)  + 1$,
where $N_s$ and $N_d$ is the number of parameters of synchrotron radiation and thermal dust emission models, respectively.
This is the caveat and weak point of the previous delta-map method because we cannot increase sensitivity by increasing $N_\nu$. We shall mitigate this point in the following sections.

Since the cleaned CMB map should contain only the CMB and the combined noise,
we can construct our likelihood as 
\begin{align}\label{eq:lh_Ichiki}
-2\ln\mathcal{L} = { \vec{m}_\mathrm{CMB} }^\T \bm{C}^{-1}  { \vec{m}_\mathrm{CMB} } +\ln|2\pi \bm{C} |,
\end{align}
where the covariance matrix $\bm{C}$ is given by~\cite{Ichiki:2018glx,Katayama:2011eh}
\begin{align}\label{eq:cov_Ichiki}
	\bm{C} = r \times \bm{C}^\mathrm{tens} +\bm{C}^\mathrm{scal} + \frac{ \bm{N}_{\nu_\mathrm{CMB}} +
		\sum_{i = \nu_2}^{\nu_{N_\nu } } \alpha_i^2  \bm{N}_{\nu_i} 
	}{ (1 + \sum_{i = \nu_2}^{\nu_{N_\nu}} \alpha_i )^2 }.
\end{align}
By minimizing this likelihood, we can determine a CMB parameter, $r$, and foreground parameters, $p_{d}^{I_d}$ and $p_{s}^{I_s}$, where $I_d \in (1, \ldots, N_d) $ and $I_s \in (1, \ldots, N_s) $. 
It is known that the determined foreground parameters are biased when we use the full likelihood function~\cite{Stompor:2008sf}.
We can avoid the bias by using the $\chi^2$ term, namely, the first term of Eq.~\eqref{eq:lh_Ichiki}.
Technically, the following iteration scheme (Algorithm~\ref{alg:Iter}) was adapted in Ref.~\cite{Ichiki:2018glx} to determine both the CMB and foreground parameters.
\begin{algorithm}
\caption{Iteration algorithm of the minimization}\label{alg:Iter}	
\begin{algorithmic}
\State Set initial values of ${p_d}^{I^d}$ and ${p_s}^{I^s}$
\State $r_\text{pre} = \infty$
\State $r_\text{out} = 1e2$
\State $-2\mathcal{L}_\text{pre} = \infty $
\State $-2\mathcal{L}_\text{out}  = 1e10$
\State Set initial values of foreground parameters
\While{$ (-2\mathcal{L}_\text{pre} + 2\mathcal{L}_\text{out}) >1e-2 $   \& 
	$(r_\text{pre}  -r_\text{out}  ) > 1e-5$
 }
\State Fix $r_\text{out}$ and minimize $\chi^2$ term against ${p_d}^{I^d}$ and ${p_s}^{I^s}$
\State Fix foreground  and minimize all the $-2\mathcal{L}$ term against $r_\text{out}$
\State Set $r_\text{pre}$ with $r_\text{out}$
\State Set $-2\mathcal{L}_\text{pre}$ with $-2\mathcal{L}_\text{out}$
\State Set $-2\mathcal{L}_\text{out}$ with the minimized $-2\mathcal{L}$
\EndWhile
\State Return $r_\text{out}$, ${p_d}^{I^d}$, and ${p_s}^{I^s}$
\end{algorithmic}
\end{algorithm}

\subsection{Extended delta-map}
In this section we introduce our method using a parametric likelihood.
We describe the details of the derivation in Appendix~\ref{sec:derivation}.

We start from Eq.~\eqref{eq:map_components}.
By subtracting the CMB and foreground terms from the observed maps $\vec{m}$, 
we can construct the likelihood of data as
\begin{align}
	-2\ln\mathcal{L} (\vec{m} | \vec{s_\mathrm{CMB}}, \sfg, \Dfg ) = (\vec{m} - \Dcmb \vec{s}_\mathrm{CMB} - \Dfg \sfg )^\T \bm{N}^{-1}(\vec{m} - \Dcmb \vec{s}_\mathrm{CMB} - \Dfg \sfg ) + \ln|2\pi \bm{N}|,
\end{align}
where $\bm{N} = \text{diag}( \bm{N_{\nu_1}}, \cdots,   \bm{N_{\nu_{N_\nu}}}) $ is a noise covariance matrix.

We use Bayes' theorem to relate the posterior distribution to the likelihood as
\begin{align}
	P( \vec{s}_\mathrm{CMB},  \sfg, \bar{p}^{I}, \Socmb, \Sfg | \vec{m}  )
	= \frac{ \mathcal{L} (\vec{m} | \vec{s}_\mathrm{CMB},  \sfg, \Dfg(\bar{p}^{I} ) ) 
	\cdot P(\vec{s}_\mathrm{CMB}, \sfg, \Dfg(\bar{p}^{I} ), \Socmb, \Sfg  ) }{ P(\vec{m} ) },
\end{align}
where $\Socmb$ and $\Sfg$ are  covariance matrices of the CMB, $\vec{s}_\mathrm{CMB}$, and foreground signal, $\sfg$, respectively, $\bar{p}^I$ represents foreground parameters of both synchrotron and thermal dust emissions, and $P(\vec{m} )$ is a normalization factor.
Here we use flat prior on $\bar{p}^I$ and have $ P(\vec{s}_\mathrm{CMB}, \sfg, \Dfg(\bar{p}^{I} ), \Socmb, \Sfg  )\propto  P(\vec{s}_\mathrm{CMB}, \sfg,  \Socmb, \Sfg  )$.
We marginalize it over the CMB signal assuming a Gaussian distribution,
\begin{align}
-2 P(\vec{s}_\mathrm{CMB}, \Socmb) = \vec{s}_\mathrm{CMB}^\T  \Socmb^{-1}\vec{s}_\mathrm{CMB} + \ln|2\pi \Socmb |,
\end{align} 
to have 
\begin{align}\begin{split}\label{eq:SNLike}
		-2 \ln P(\bar{p}^{I}, \sfg, \Sfg, \Socmb |\vec{m} ) &= 
		\left(  \vec{m} - \Dfg  \sfg  \right)^\T \left( \Scmb  + \bm{N} \right)^{-1} 
		\left( \vec{m} - \Dfg \sfg \right)  
		\\&	+ 
		\ln \left| 2  \pi \left(  \Scmb + \bm{N}  \right) \right| 
		-2 \ln P \left( \sfg ,  \Sfg  \right) +  \mathrm{const.}\, ,
\end{split}\end{align}
where $\Scmb =\Dcmb \Socmb \Dcmb^\T$.
This is Eq.~(58) in Ref.~\cite{Ichiki:2018glx}.

From here, we deal with $P \left( \sfg ,  \Sfg  \right)$ term.
Our strategy is to marginalize foreground probability function assuming its mean is zero, $\sfg = \vec{0}$ and its covariance is ``vague'',  $\Sfg^{-1} \rightarrow \bm{O}$.
We follow the methodology described in Sect.~$2$ of  Ref.~\cite{Rasmussen:2006} to deal with the ``vague'' foreground covariance matrix.
Following the method, we obtain 
\begin{align}\begin{split}\label{eq:SNLikeFinal}
		-2\ln P(\bar{p}^{I}, \Socmb | \vec{m} ) &=
		- \vec{m}^\T \bm{N}^{-1} \Dcmb \bm{A}^{-1} {\Dcmb}^\T \bm{N}^{-1}\vec{m}
		\\&-\vec{M}\bm{H}\vec{M} - \vec{M}\bm{H}\bm{B}^{-1}\bm{H}\vec{M}
		\\& + \ln |\Socmb|  
		+ \ln| { \Dfg }^\T \bm{N}^{-1} \Dfg  |+ \ln|\bm{B}|+  \mathrm{const.}\, ,
\end{split}\end{align}
where
\begin{align}
	\bm{A} &= ( \Socmb^{-1}   + \sum_{j =1}^{N+1} \bm{N}_{\nu_j}^{-1} )~,\\
	\bm{H} &= \bm{N}^{-1} \Dfg \left[  {\Dfg}^\T \bm{N}^{-1} \Dfg\right]^{-1} \Dfg^\T \bm{N}^{-1}~,\\
	\bm{B} &= \bm{A} - \Dcmb^\T \bm{H} \Dcmb~,\\
	\vec{M} &=\vec{m} - \Dcmb \bm{A}^{-1}  \Dcmb^\T \bm{N}^{-1} \vec{m}~,
\end{align}
and put the noise terms, $\vec{m}^\T  \bm{N}^{-1} \vec{m} +\ln|\bm{N} |$, into the $\text{const.}$ term.
For details of the derivation of Eq.~\eqref{eq:SNLikeFinal}, see Appendix~\ref{sec:derivation}. To determine the CMB parameter ($r$) and foreground parameters ($\bar{p}^{I}$), we follow the same procedure, namely Algorithm~\ref{alg:Iter}, as in Ref.~\cite{Ichiki:2018glx}.
We set the initial values of all parameters as,
$r =0.5$, $\beta_s = -3.0$, $\beta_d = 1.5$, and $T_d =\SI{20.1}{K}$.
We set lower bound on $r \geq 0.0$ so that the CMB covariance matrix is not singular and set boundaries on foreground parameters as,
$\beta_s \in (-10.0, -0.01)$, 
$\beta_d \in (0.1, 10.0)$,
and $T_d \in (5.0, 40.0)$.

We obtain the ``extended delta-map'' likelihood as Eq.~\eqref{eq:SNLikeFinal}, which allows us to determine CMB and foreground parameters without the band number constraint.
This likelihood has another benefit.
Because we use ``matrix inversion lemma'' (Appendix~\ref{sec:lemma}) in the derivation, 
we can reduce the computational cost in using the Cholesky solver for $\Scmb+\bm{N}$ in Eq.~\eqref{eq:SNLike}, which is positive definite symmetric matrix with a dimension of $2N_\text{pix}N_{\nu}$.
In Eq.~\eqref{eq:SNLikeFinal}, on the contrary,
we only need to use the Cholesky solver for some positive definite symmetric matrices, e.g., $\bm{B}$, with a smaller size of $2N_\text{pix}\left(N_d + N_s + 3 \right) $.

Because the form of Eq.~\eqref{eq:SNLikeFinal} looks so different from that of Eq.~\eqref{eq:lh_Ichiki},
one may suspect that they are totally different methods.
We find that estimate values of the parameters with this likelihood are equivalent to those estimated from the previous delta-map method in the case where $N_\nu= \left( N_d(=1) + 1\right) + 1$.
We describe the comparison in Appendix~\ref{sec:Comparison}. 

All of the numerical codes for the calculations above have been implemented in a \texttt{extended-deltamap} GitHub repository~\footnote{\url{https://github.com/YutoMinami/extended-deltamap}}.
\section{Models and simulations\label{sec:sim}}

We use simulations to validate our methodology.
We use the ``PySM'' package~\cite{Thorne:2016ifb} to produce polarized synchrotron and thermal dust emission maps with direction-dependent spectral parameters.
For the synchrotron map, we use the power-law synchrotron model, "s1", which is based on the $Q$ and $U$ maps from WMAP-9~\cite{Bennett_2013} and the spectral index map from Ref.~\cite{Miville-Deschenes:2008lza}.
For the thermal dust emission map, we use the one-component modified black-body (MBB) model, "d1"~\cite{Planck:2015mvg}, and the two-component MBB model, "d4"~\cite{2015ApJ...798...88M}, both of which are based on the Planck HFI results.
Note that the results used the intensity map in addition to $Q$ and $U$ polarization maps, and the model adopted common foreground parameters $\beta_d$ and $T_d$ for $Q$ and $U$ Stokes parameters at each sky pixel.
Cosmological CMB maps are generated using \texttt{synfast} function of HEALPix package~\cite{Gorski:2004by} from the power spectra calculated using CAMB~\cite{Lewis:2000}
with the Planck 2018 cosmological parameters for ``TT,TE,EE$+$lowE$+$lensing''~\cite{Aghanim:2018eyx}:
$\Omega_bh^2=0.02237$, $\Omega_ch^2=0.1200$,
$h=0.6736$, $\tau=0.0544$, $A_s=2.100\times 10^{-9}$, and $n_s=0.9649$.
We generate CMB maps with some values of the tensor-to-scalar ratio, $r$.

We use experimental parameters such as frequency bands and angular resolutions similar to the LiteBIRD mission~\cite{Hazumi2019} (Table~\ref{tab:LBspec}).
For the input noise, we assume white noise with standard deviation of $\sigma_\mathrm{N} = (\pi/10800)(w_\mathrm{p}^{-1/2}/\mu\mathrm{ K^{'} })~\mu \mathrm{K~str^{-1/2}} $~\cite{Katayama:2011eh}, where 
we use the ``Polarization sensitivity'' column of Table~\ref{tab:LBspec} for the values of $w_\mathrm{p}^{-1/2}$.

We incorporate the beam smearing effect from the finite angular resolution, whose full-width-half-maximum (FWHM) values are given by the ``Beam size in FWHM'' column of Table~\ref{tab:LBspec},
by multiplying the spherical harmonics coefficients of the CMB and foreground maps by the corresponding Gaussian beam transfer function for individual frequencies.
In this paper, we use the map resolution parameter of $N_\text{side} = 4$ and set $\ell_\text{max} = 2N_\text{side}$.
To ensure that the maps are limited to low resolution, 
we follow the method described in Ref.~\cite{Eriksen:2006xr};
specifically, 
we de-convolve each frequency map and re-convolve all the maps with Gaussian \SI{2200}{arcmin} beam,
which is $2.5$ times the pixel size of $N_\text{side}=4$.
Even though our method can be applied to higher resolution maps, we choose $N_\text{side}=4$ (or $\ell_\text{max} = 8$) which covers the so-called reionization bump.
Since we do not apply any delensing scheme,
we cannot effectively increase the sensitivity to $r$ because of the cosmic variance of lensing $B$-mode.
If we can increase $\ell_\text{max}$ further to cover the so-called recombination bump, we could increase the sensitivity.
However, such high-resolution analysis is limited by our computational resources.

In Sect.~\ref{sec:ModHigh}, we will show the case in which the frequencies of the two highest frequency bands are increased.
For that we show the replaced parameters in brackets.
The replaced polarization sensitivity and beam size in FWHM are extrapolated using power-law function.

Since our method approximates the directional dependence of foreground parameters up to the first order,
we need to mask the very bright Galactic plane.
We follow Ref.~\cite{Ichiki:2018glx} and use the ``P06 mask'' by WMAP polarization analysis, whose $f_\text{sky}=0.56$.
Since our method uses pixel space maps, 
we do not apply any apodization to the mask.

\begin{table}
	\centering
	\begin{tabular}{c c c}
		\toprule
		Frequency (GHz) & Polarization sensitivity ($\mathrm{\mu K^{'}}$) & Beam size in FWHM (arcmin) \\ 
		\midrule
		40 & 37.5 & 69 \\
		50 & 24.0 & 56 \\
		60 & 19.9 & 48 \\
		68 & 16.2 & 43 \\
		78 & 13.5 & 39 \\
		89 & 11.7 & 35 \\
		100 & 9.2 & 29 \\
		119 & 7.6 & 25 \\
		140 & 5.9 & 23 \\
		166 & 6.5 & 21 \\
		195 & 5.8 & 20 \\
		235 & 7.7 & 19 \\
		280 & 13.2 & 24 \\
		337 (500) & 19.5 (67.8) & 20 (13.7)\\
		402 (600)& 37.5 (113.5) & 17 (11.5)\\
		\bottomrule
	\end{tabular}
	\caption{Polarization sensitivity and beam size of the telescopes similar to the LiteBIRD mission~\cite{Hazumi2019}.
	The parameters written in brackets are for the modified high-frequency model used in Sect.~\ref{sec:ModHigh}.
	}\label{tab:LBspec}
\end{table}

\section{Results\label{sec:results}}
We apply our new method to simulated CMB maps with foregrounds.
In the following sections, we show the results of estimating tensor-to-scalar ratio $r$ with CMB + synchrotron (Sect.~\ref{sec:results_synch}), CMB + 1MBB dust (Sect.~\ref{sec:results_dust}), and CMB + 1MBB dust + synchrotron (Sect.~\ref{sec:results_dust_synch}) maps.
Recently, it has been shown that shifting the observation bands toward higher frequencies can improve the determination of foreground parameters~\cite{LiteBIRD:2022cnt}.
Thus, we show the results of estimating tensor-to-scalar ratio $r$,
with the two highest observation frequencies given in Table~\ref{tab:LBspec}
shifted even higher,
in Sect.~\ref{sec:ModHigh}.
Finally, we show the results with the two-component dust model as a case of mismodeling of the foreground emissions.

We summarize all the estimated $r_\text{out}$ values for all the setups in Table~\ref{tab:results}.
\begin{table}
	\centering
\caption{\label{tab:results}
Summary of our analysis.
Estimated $r_\text{out}$ value is from $50$th percentile.
For $r_\text{in}=\SI{1.0e-2}{}$, we use ($16$th,$84$th) percentiles for its uncertainties.
For null $r_\text{in}$, we use $95\%$ C.L. as the upper bounds.
``(High)'' indicates the result with ``modified high-frequency model'' where we increase the frequencies of the highest two frequency bands.
}
\begin{tabular}{cccccc}
	\toprule
	Number of bands& Input synchrotron& Input dust &  $T_d$ prior &$r_\text{in}\times 10^{-2}$&$r_\text{out} \times 10^{-2}$  \\
	\midrule
	$ 3$ & s1 & -  &   -  &$\SI{1.0e0}{}$&$1.00_{-0.33}^{+0.18}$\\
	$ 9$ & s1 & -  &   -  &$\SI{1.0}{}$&$0.99_{-0.28}^{+0.15}$\\
	$ 3$ & s1 & -  &   -  &$\SI{0.0}{}$&$ < 0.12$ ($95\%$C.L)\\
	$ 9$ & s1 & -  &   -  &$\SI{0.0}{}$&$ < 0.07$ ($95\%$C.L)\\
	$ 4$ & -  & d1 & flat &$\SI{1.0}{}$& $0.95_{-0.95}^{+1.43}$\\
	$ 9$ & -  & d1 & flat &$\SI{1.0}{}$&$1.13_{-0.49}^{+0.61}$\\
	$15$ & s1 & d1 & flat &\SI{1.0}{}&${0.96}_{-0.96}^{+0.49} $\\
	$15$ & s1 & d1 & $1\sigma$ &\SI{1.0}{}&${1.08}_{-0.90}^{+0.52}$\\
	$15$ & s1 & d1 & $10^{-5}\sigma$ &\SI{1.0}{}&${1.18}_{-0.83}^{+0.50}$\\
	$15$ & s1 & d1 & flat &\SI{0.0}{}&$ < 1.15$ ($95\%$C.L)\\
	$15$ & s1 & d1 & $1\sigma$ &\SI{0.0}{}&$ < 1.26$ ($95\%$C.L)\\
	$15$ (High) & s1 & d1 & flat &\SI{1.0}{}&${1.40}_{-0.48}^{+0.63}$\\
	$15$ (High) & s1 & d1 & $1\sigma$ &\SI{1.0}{}&${1.41}_{-0.42}^{+0.56}$\\
	$15$ & s1 & d4 & flat &\SI{1.0}{}&${2.00}_{-0.85}^{+1.09}$\\
	$15$ & s1 & d4 & $1\sigma$ &\SI{0.0}{}&${2.19}_{-0.93}^{+1.26}$\\	
	\bottomrule
\end{tabular}
\end{table}

\subsection{Estimation with synchrotron radiation foreground only}\label{sec:results_synch}
\begin{figure}
	\centering
	\begin{minipage}{0.48\textwidth}
	\centering
	\includegraphics[width=1.0\columnwidth]{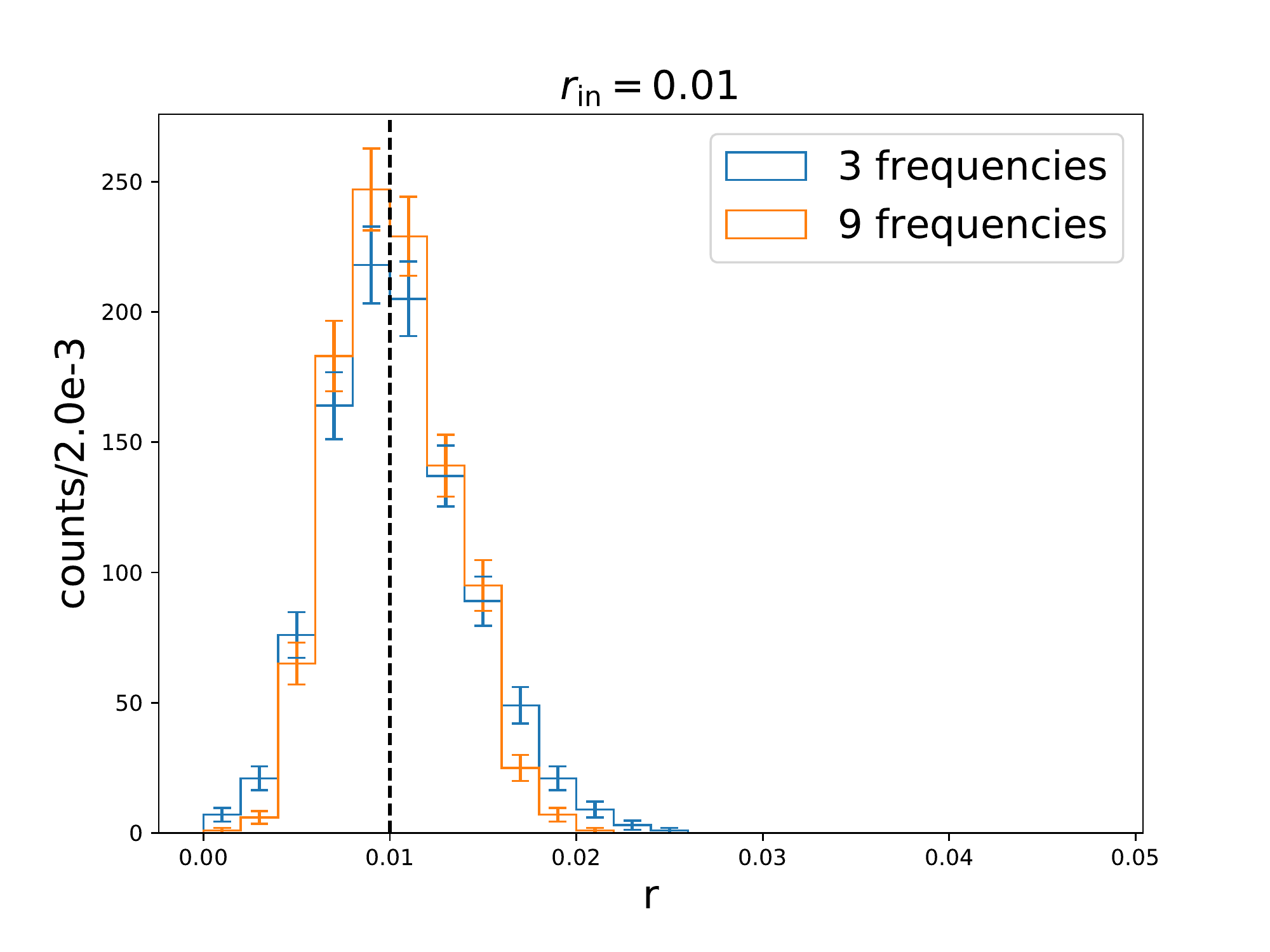}
	\end{minipage}
    \hfill
	\begin{minipage}{0.48\textwidth}
	\centering
    \includegraphics[width=1.0\columnwidth]{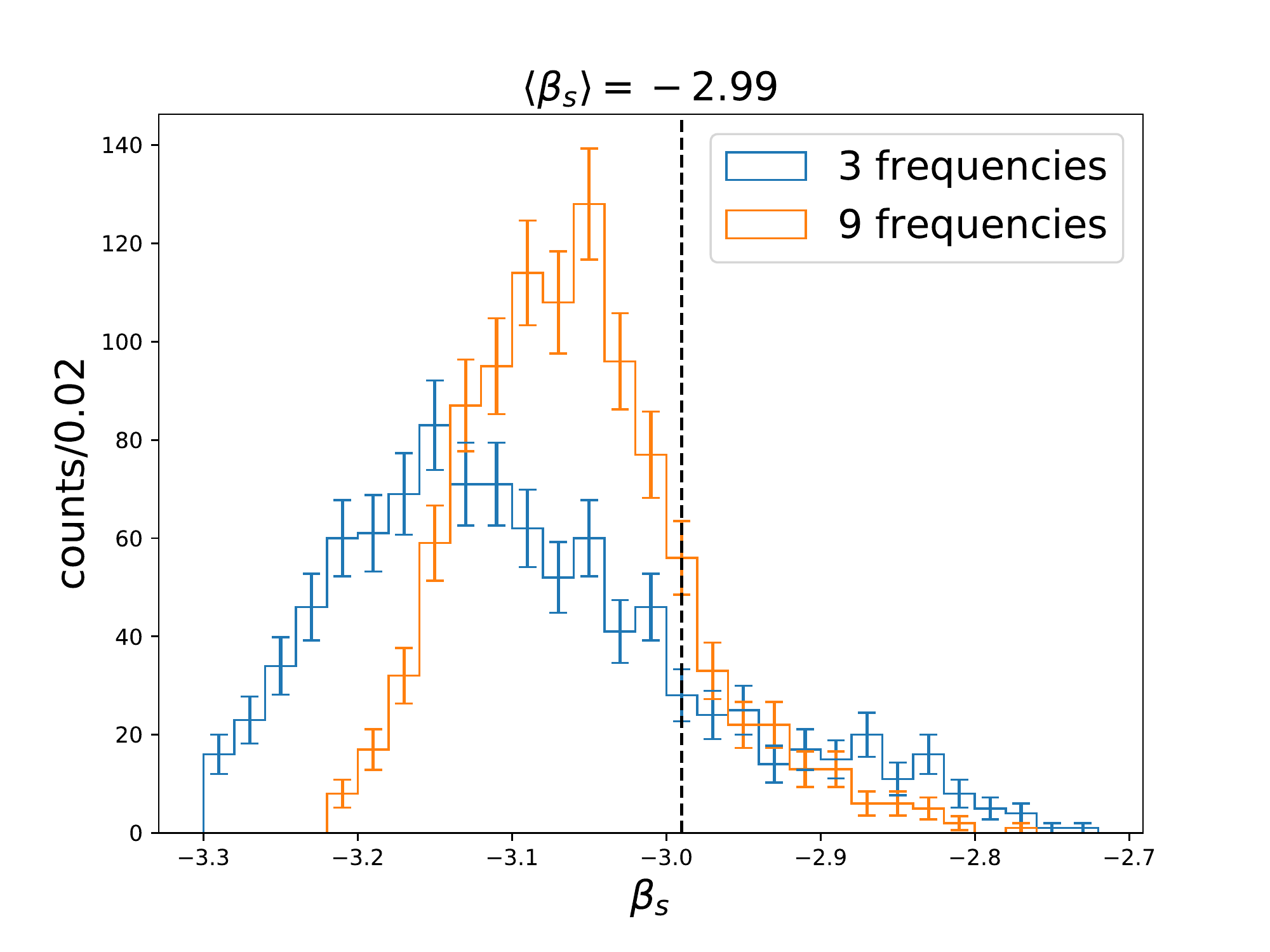}
    \end{minipage}
    \caption{\label{fig:Synch_var_LTD17}
	Estimated $r_\text{out}$ (left) and $\beta_s$ (right) from $1000$ realizations.
	Blue histograms show the estimates with three bands and orange histograms show the estimates with the nine bands.
	}
\end{figure}

In this section we consider the case where only synchrotron radiation is the foreground source.
Because we use a power-law synchrotron radiation model, we need at least three observation bands to fit our parametric model.
We first estimate tensor-to-scalar ratio,  $r$, and synchrotron spectrum index, $\beta_s$, with three exact bands $\nu\in(40,60,140)\,\si{GHz}$ against input $r_\text{in}=0.01$.
We show the histograms of the  estimated $r$ and $\beta_s$ as blue boxes of the left and right panels of  Fig.~\ref{fig:Synch_var_LTD17} from $1000$ realizations, respectively.
The black vertical line shows the naive average of the $\beta_s$ of PySM input in the unmasked sky area.

We next estimate $r$ and $\beta_s$ with nine bands $\nu\in(40,50,60,68,78,89,100,119,140)\,\si{GHz}$, and
show the histograms of the estimated $r$ and $\beta_s$ in the left and right panels of Fig.~\ref{fig:Synch_var_LTD17} in orange, respectively.
This is one of the main results of this paper showing that we are free from the constraint of the number of bands that existed in the previous delta-map method.
We find that $r$ and $\beta_s$ are determined better by increasing the number of the observation bands.

\begin{figure}
	\centering
\includegraphics[width= 0.48\columnwidth]{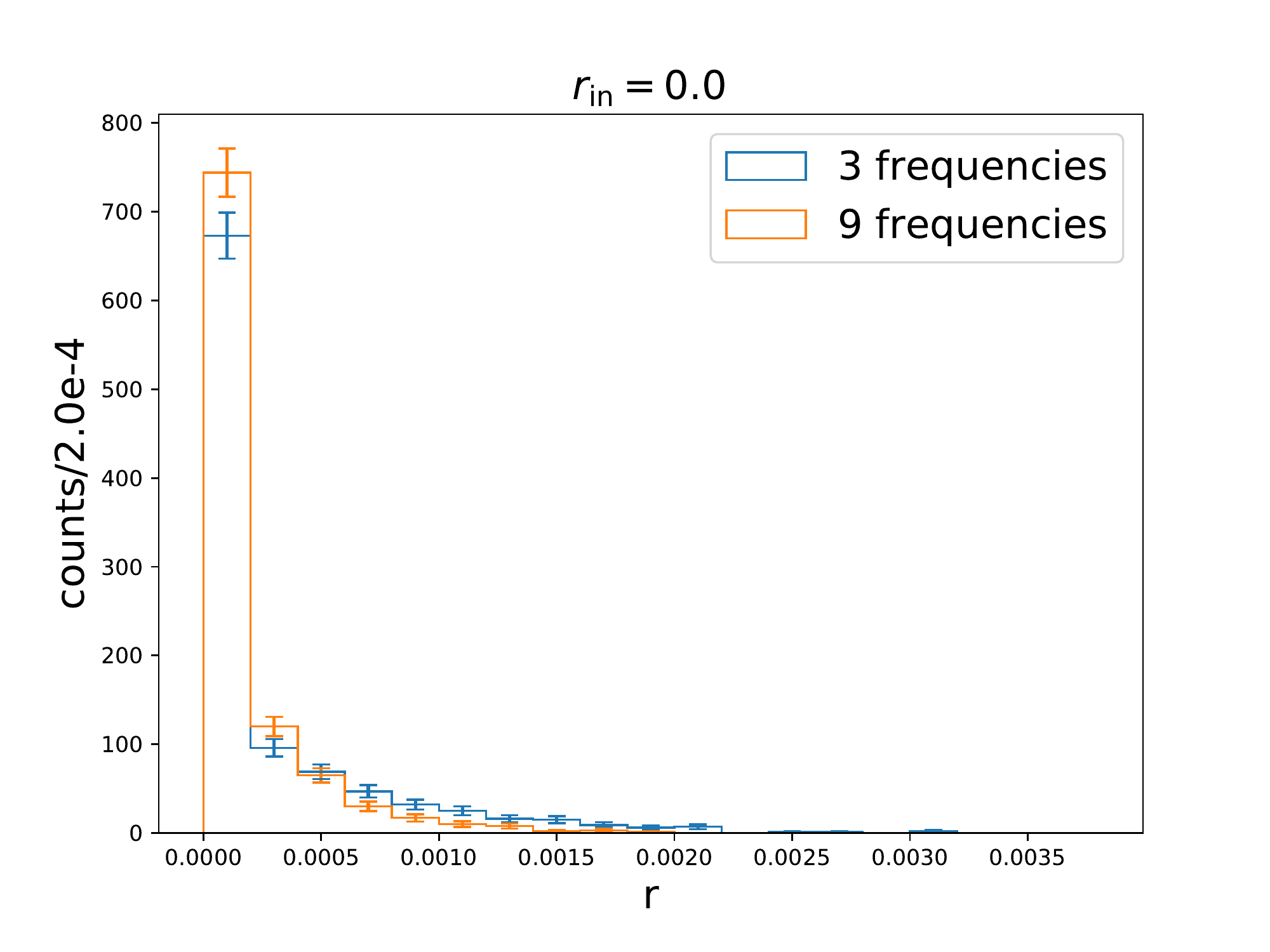}
\caption{\label{fig:Synch_var_LTD17_r0e0}
	Estimated $r_\text{out}$ from $1000$ realizations for null $r_\text{in}$.
	Blue histograms show the estimated results with three bands and orange histograms show those with nine bands. 
}
\end{figure}

We show the results of the estimation of $r_\text{out}$ for the case in which $r_\text{in}$ is null in Fig.~\ref{fig:Synch_var_LTD17_r0e0}.
We find that $r$ is constrained to $r_\text{ out} < \SI{0.12e-2}{}$ and $r_\text{out}< \SI{0.07e-2}{}$ using three and nine bands, respectively.

\subsection{Estimation with thermal dust emission foreground only}\label{sec:results_dust}
Next, we consider the case where only the thermal dust emission is the foreground source.
We choose one-component modified black body (1MBB) model as the foreground dust model, for which we need at least four bands to fit our parametric model, and we set $r_\text{in}$ to be $0.01$.

\begin{figure}
	\centering
	\includegraphics[width= 0.48\columnwidth]{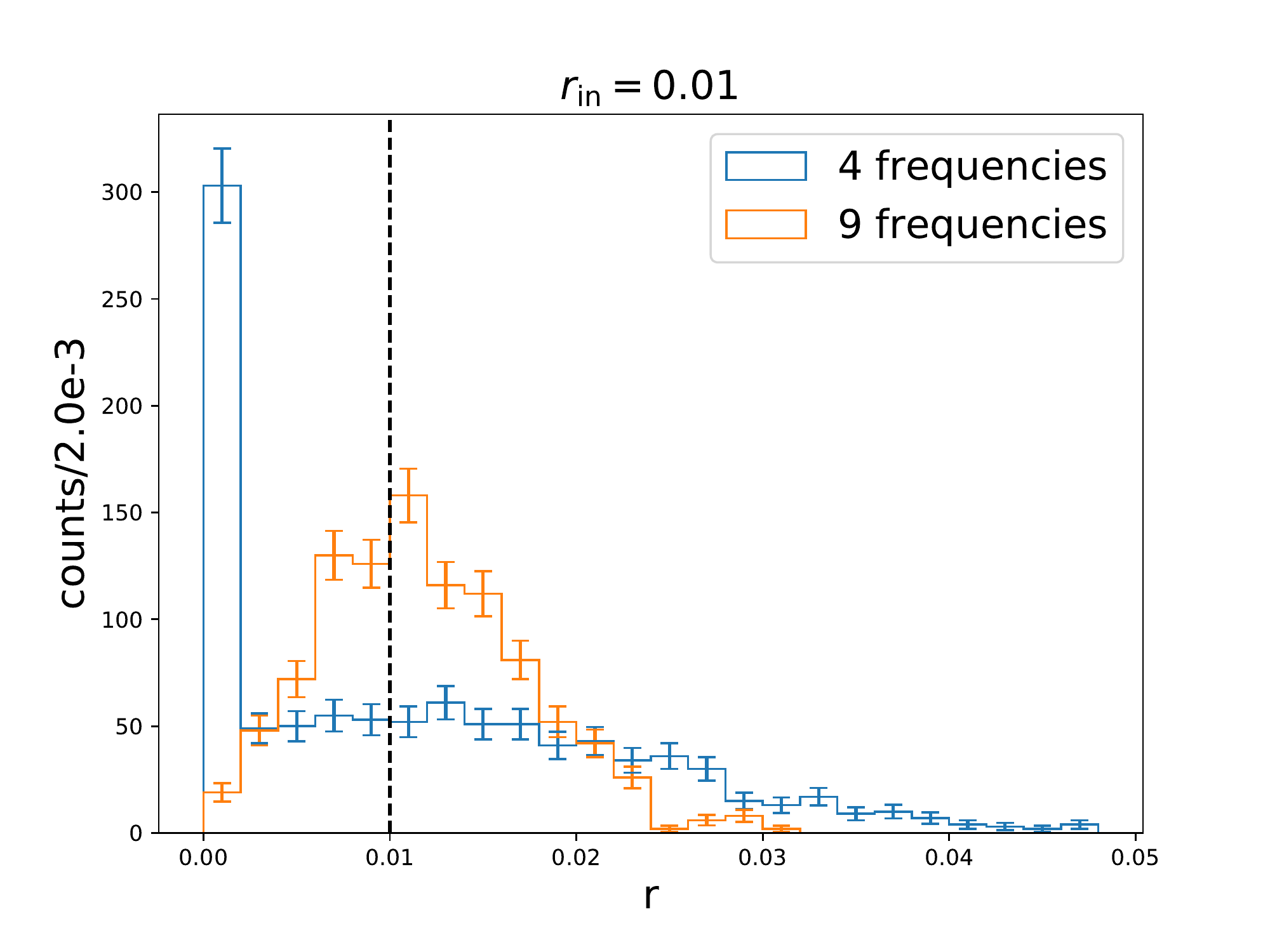}
	\caption{
		Histograms of estimated $r$ from 1000 realization samples.
		Blue histogram shows the estimated $r$ with four bands.
		Orange histogram shows the estimated $r$ with nine bands.
		\label{fig:Dust_var_LTD17_r1e-2}
	}
\end{figure}

We first estimate the tensor-to-scalar ratio parameter $r$ and foreground parameters, $T_d$ and $\beta_d$, with the four exact bands, $\nu\in(140, 235, 280, 402)\,\si{GHz}$.
We show the histograms of the estimated $r_\text{out}$ from $1000$ realizations in Fig.~\ref{fig:Dust_var_LTD17_r1e-2}.
We find that uncertainty on $r_\text{out}$ is large and the lowest bin is dominant, which reflects the fact that we impose an $r \geq 0$ prior on $r$ so that $\Socmb$ is a positive definite matrix.

\begin{figure}
	\centering
	\begin{minipage}{0.48 \textwidth}
	\includegraphics[width= 1.0\columnwidth]{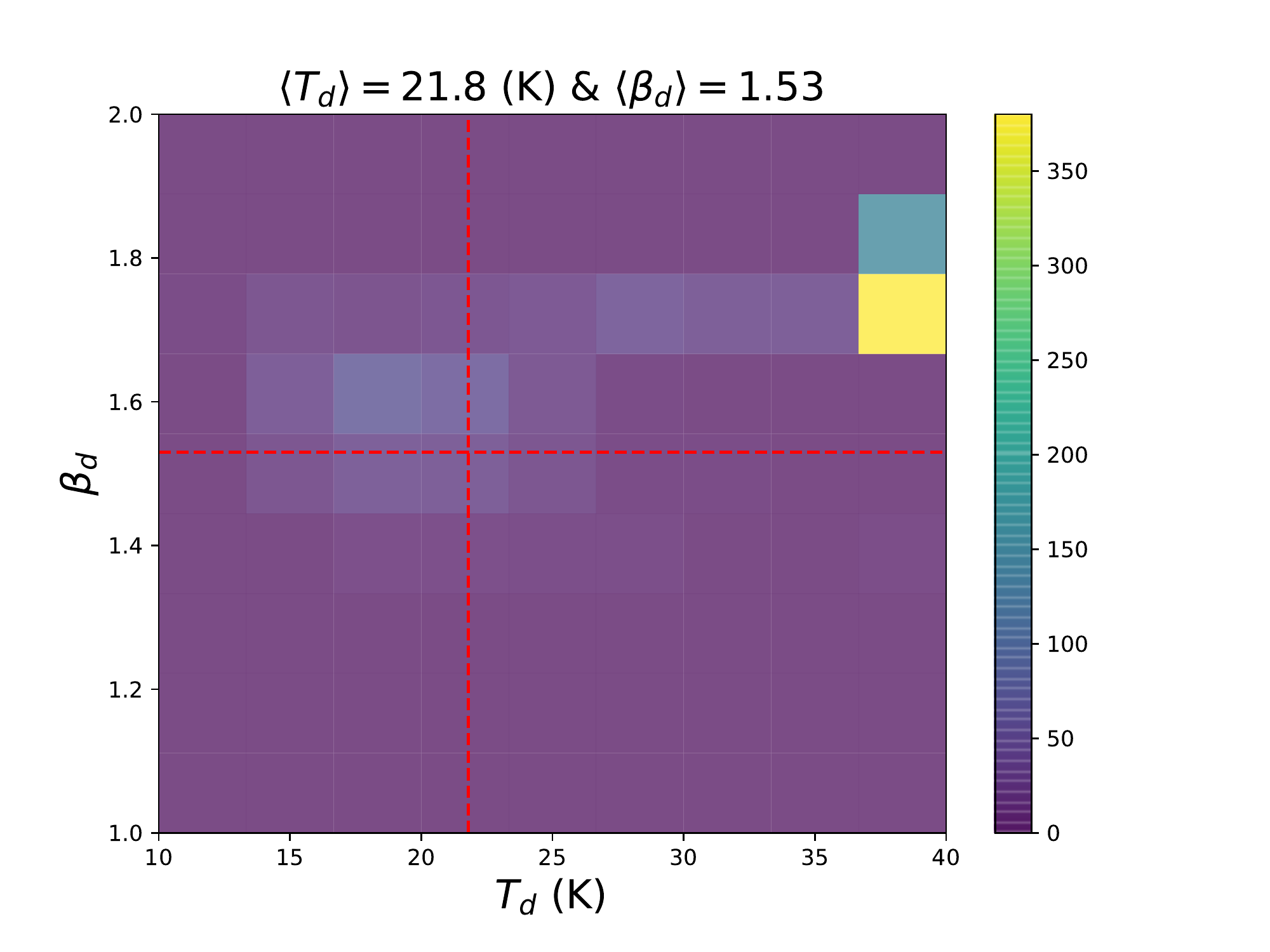}
	\end{minipage}
\hfill
	\begin{minipage}{0.48\textwidth}
	\includegraphics[width= 1.0\columnwidth]{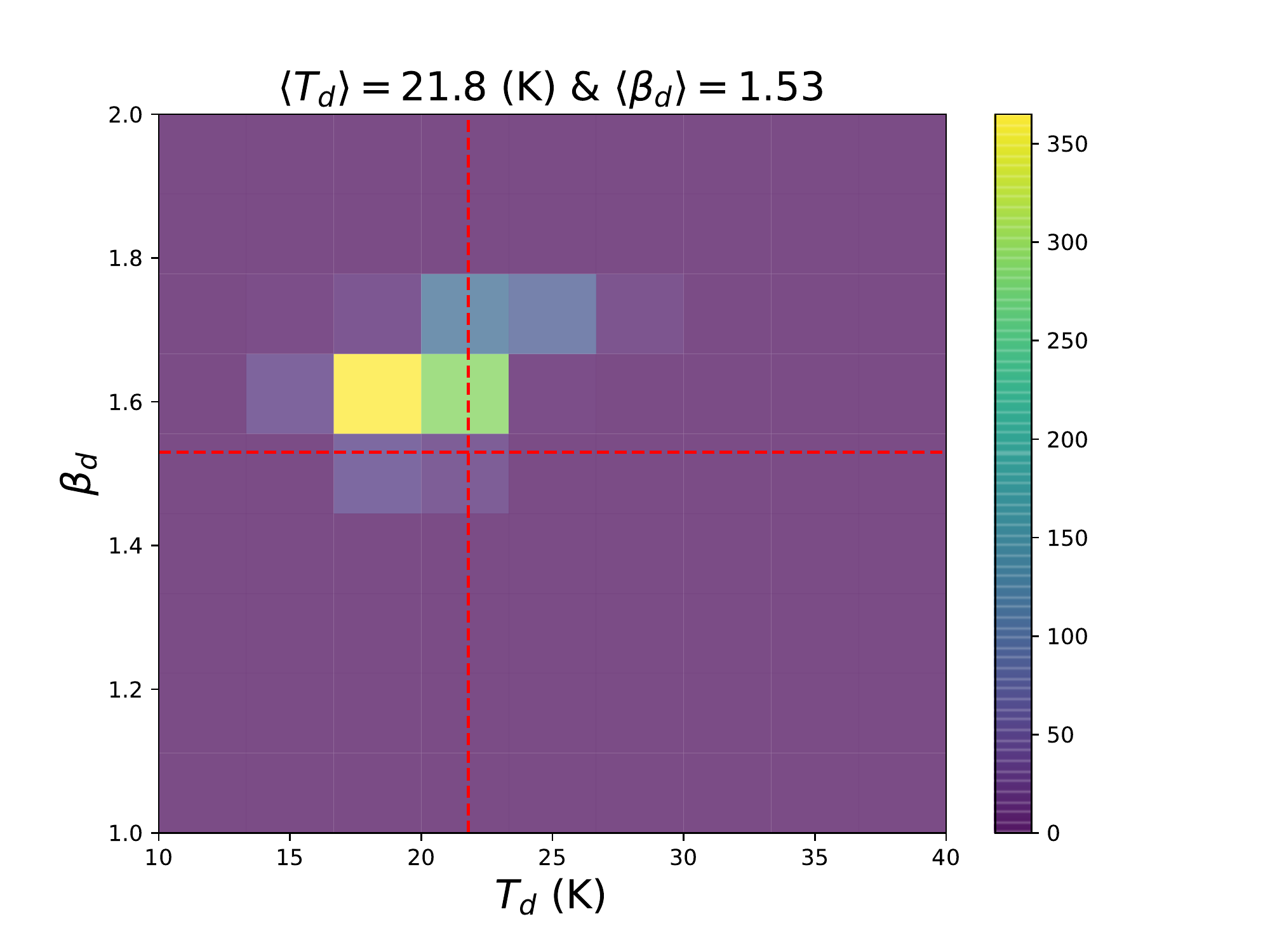}
\end{minipage}
\caption{\label{fig:Dust_var_4and9freq_Td_vs_betad}
		2D histograms of estimated dust parameters, $\beta_d$ vs $T_d$ with four bands (left) and nine bands (right). 
		}
\end{figure}

We show 2D histograms of the estimates of $T_d$ and $\beta_d$ in the left panel of Fig.~\ref{fig:Dust_var_4and9freq_Td_vs_betad}.
We find that foreground parameters $T_d$ and $\beta_d$ are not precisely determined with the four bands.
The undetermined foreground parameters cause the large error in estimating $r$.

This can be mitigated by increasing the number of frequency bands.
Next we show the estimated $r_\text{out}$ with nine bands, $\nu\in(100,119,140,166,195,235,280,337,402)\,\si{GHz}$, in the orange histogram of Fig.~\ref{fig:Dust_var_LTD17_r1e-2}, 
and the estimated $T_d$ and $\beta_d$ with nine bands using the 2D histogram in the right panel of Fig.~\ref{fig:Dust_var_4and9freq_Td_vs_betad}.
We find that we can determine foreground parameters well using the nine bands, and thereby we can determine $r_\text{out}$ more precisely.


\subsection{\label{sec:results_dust_synch}Estimation with thermal dust and synchrotron foreground emissions}
Finally, we consider one-component modified black body dust and power-law synchrotron as our foreground models, and estimate the CMB parameter, $r$, and the foreground parameters, $T_d$, $\beta_d$, and $\beta_s$.
We use all the $15$ bands in Table~\ref{tab:LBspec}.
Because we found that the number of bands is not sufficient to determine $T_d$, we impose some priors, $T_d = 21.8 \pm 1\sigma\,$K and $\pm \SI{1.0e-5}{}\sigma$\,K, on $T_d$,
where $\sigma$ is the standard deviation of the dust temperature measured by Planck~\cite{Planck:2015mvg}.

\begin{figure}
\centering
    \begin{minipage}{0.45\textwidth}
	\includegraphics[width = 1.0\columnwidth]{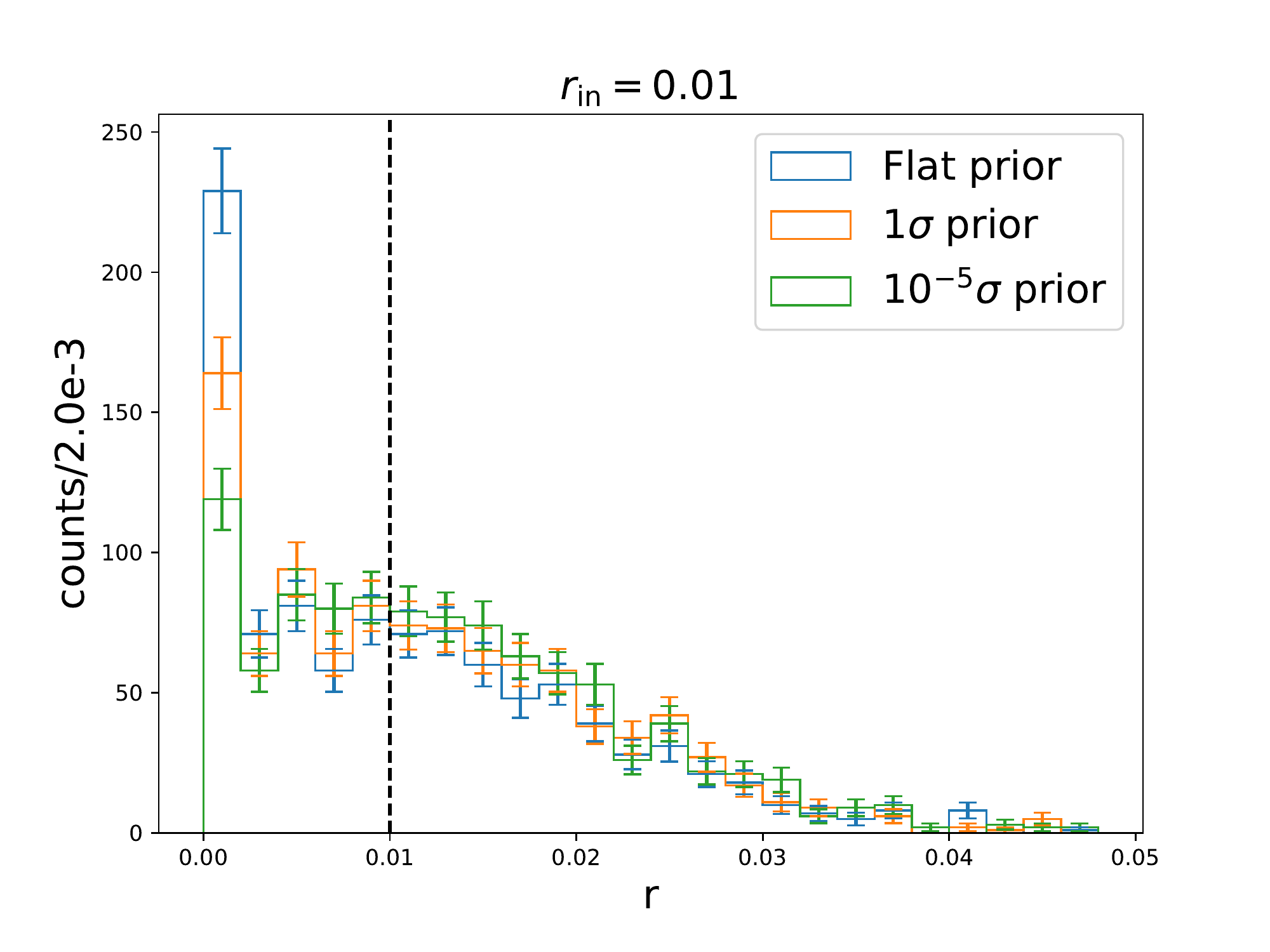}
    \end{minipage}
    \hfill
    \begin{minipage}{0.45\textwidth}
	\centering
	\includegraphics[width = 1.0\columnwidth]{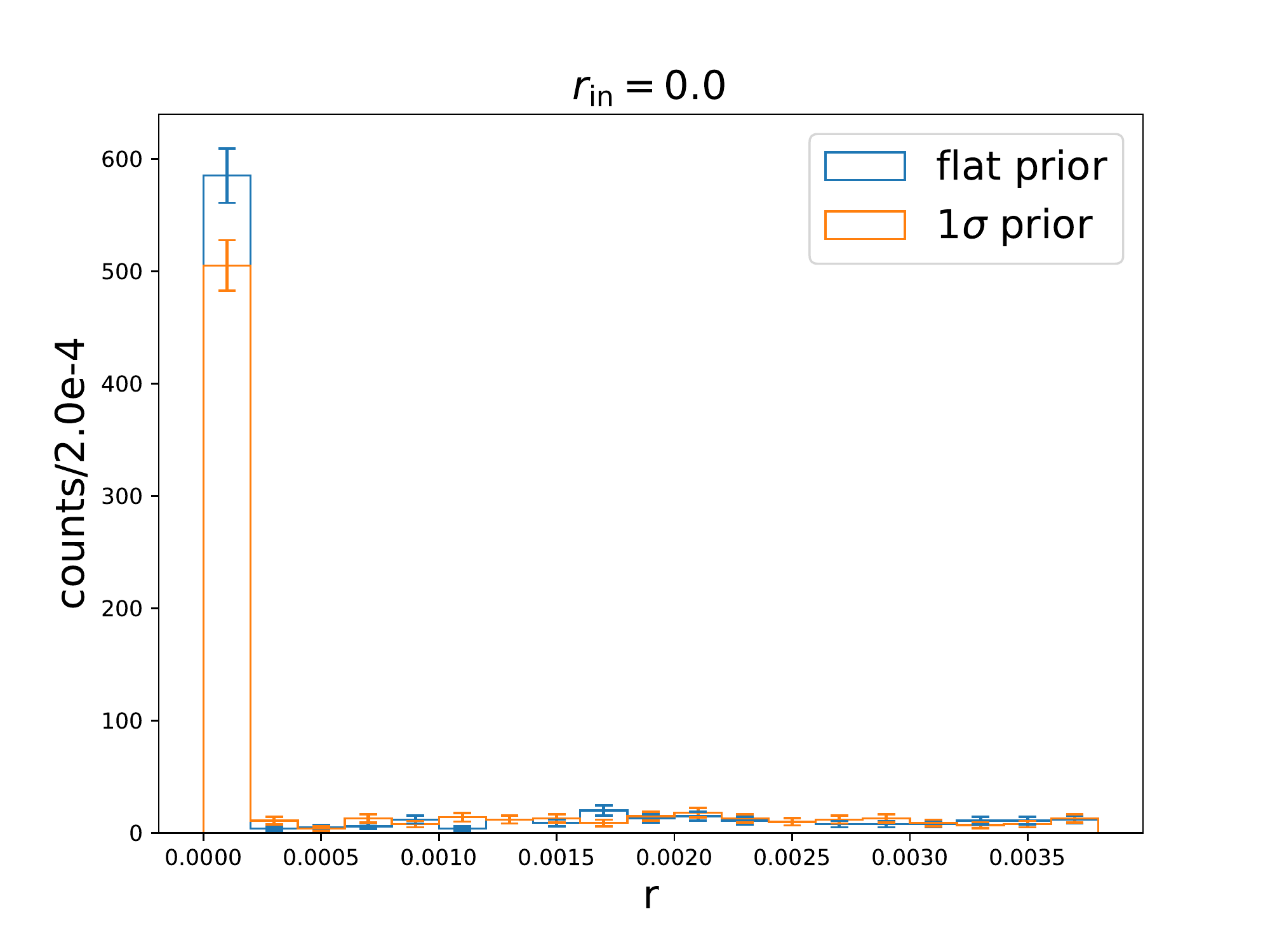}
	\end{minipage}
	\caption{\label{fig:Dust_Synch_var_LTD17_15freq}
	Histograms of estimated $r_\text{out}$ from 1000 samples.
	(Left) $r_\text{out}$ histograms for input $r_\text{in} = 0.01$.
	Blue, orange, and green histograms show the estimated $r_\text{out}$ with flat, $1\sigma$, and $10^{-5}\sigma$ priors on $T_d$, respectively.
	(Right) $r_\text{out}$ histograms for input $r_\text{in} = 0.0$. 
	Blue and orange histograms show the estimated $r_\text{out}$ with flat and $1\sigma$ priors on $T_d$, respectively.	
	}
\end{figure}

The estimated $r_\text{out}$ are shown in the left panel of Fig.~\ref{fig:Dust_Synch_var_LTD17_15freq}.
We find that stronger constraints on $T_d$ lead to a more precise estimate of $r_\text{out}$.
However, the $50$th percentile is slightly biased to positive, as shown in Table~\ref{tab:results}.
We can see this bias in the estimation of $r$ for null $r_\text{in}$, as shown in the right panel of Fig.~\ref{fig:Dust_Synch_var_LTD17_15freq}, which shows $r_\text{out}$ is slightly biased to positive and the upper bound with $95\%$ C.L becomes larger (Table~\ref{tab:results}).

\subsection{\label{sec:ModHigh} Modified high-frequency case }
We next see the results with the modified high-frequency model, which increase frequencies of the high-frequency bands of telescopes.

\begin{figure}
	\centering
\begin{minipage}{0.45\textwidth}
		\centering
		\includegraphics[width=\columnwidth]{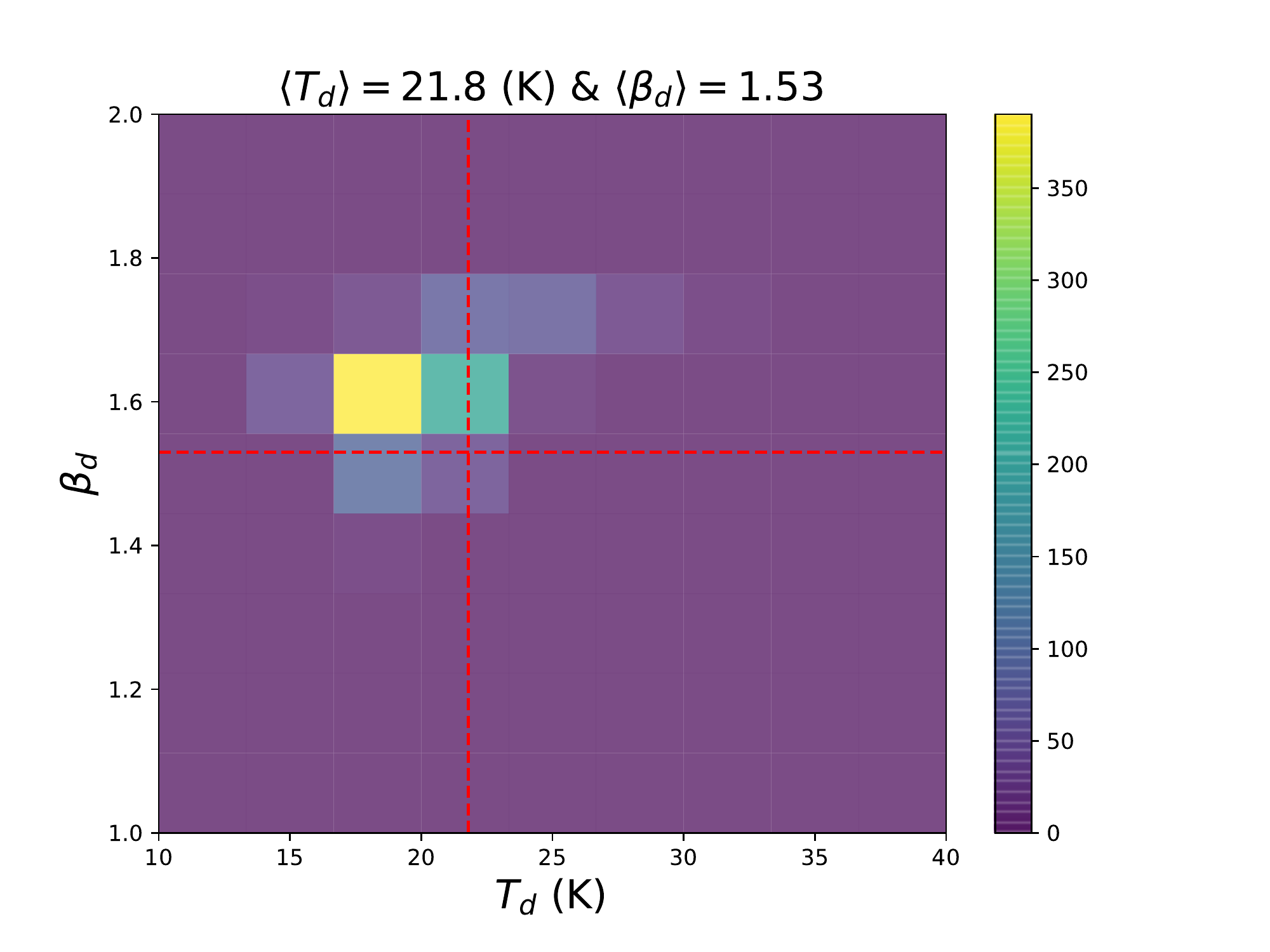}
\end{minipage}	
    \hfill
	\begin{minipage}{0.45\textwidth}
		\centering
		\includegraphics[width=\columnwidth]{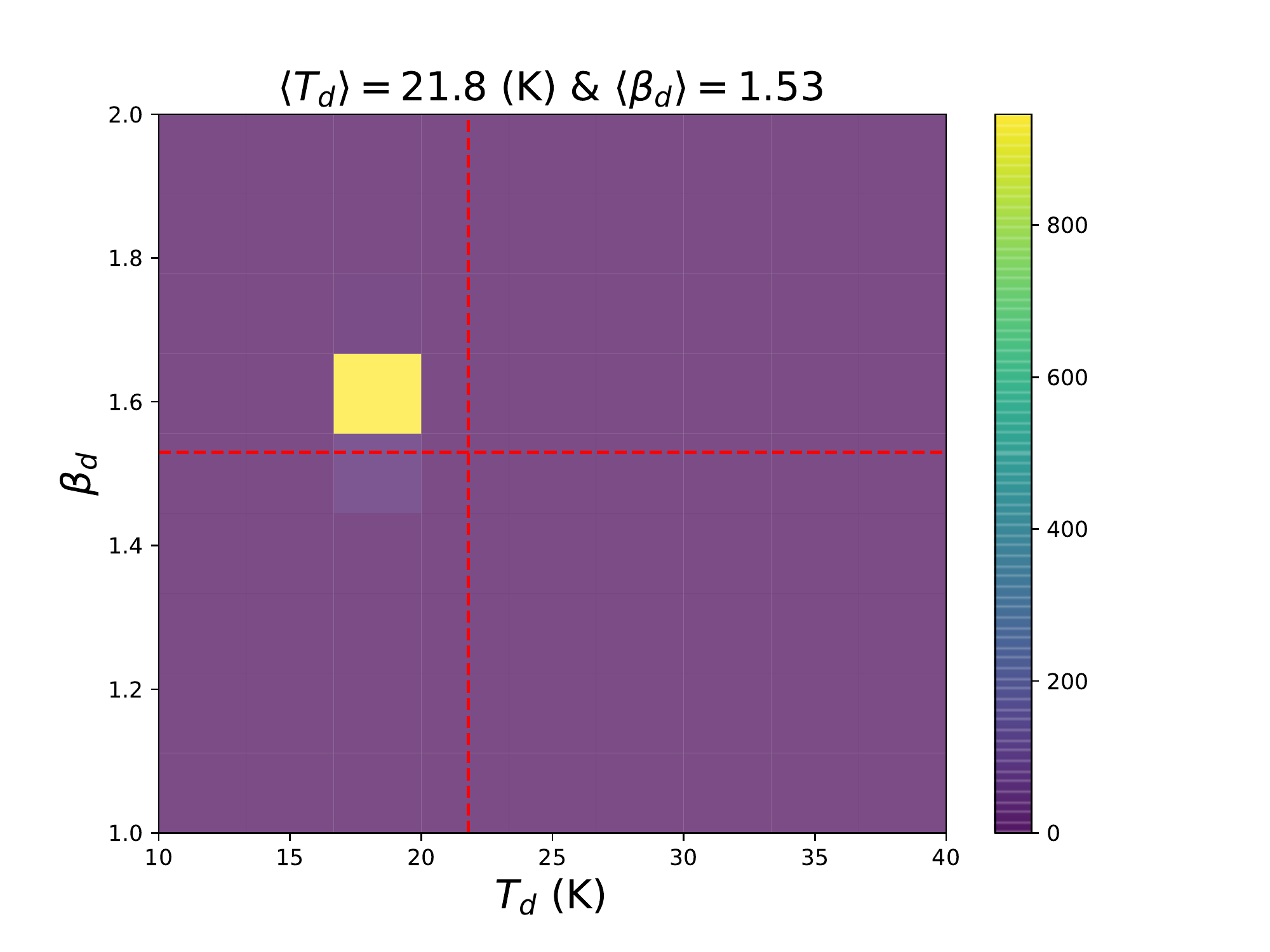}
	\end{minipage}
\caption{\label{fig:Dust_Synch_var_LTD_high_15freq_r1e-2_Td_vs_betad_comparison}
2D histograms of $T_d$ vs $\beta_d$ with $15$ normal bands (left) and $15$ modified high-frequency bands (right).}
\end{figure}

To see the determination of foreground parameters,
we show 2D histograms of $\beta_d$ and $T_d$ for $15$ normal bands and $15$ modified bands in the left and right panels of Fig.~\ref{fig:Dust_Synch_var_LTD_high_15freq_r1e-2_Td_vs_betad_comparison}, respectively.
 We find that we can determine foreground parameters precisely in the case of modified high-frequency model.
 
\begin{figure}
	\centering
	\includegraphics[width= 0.48\columnwidth]{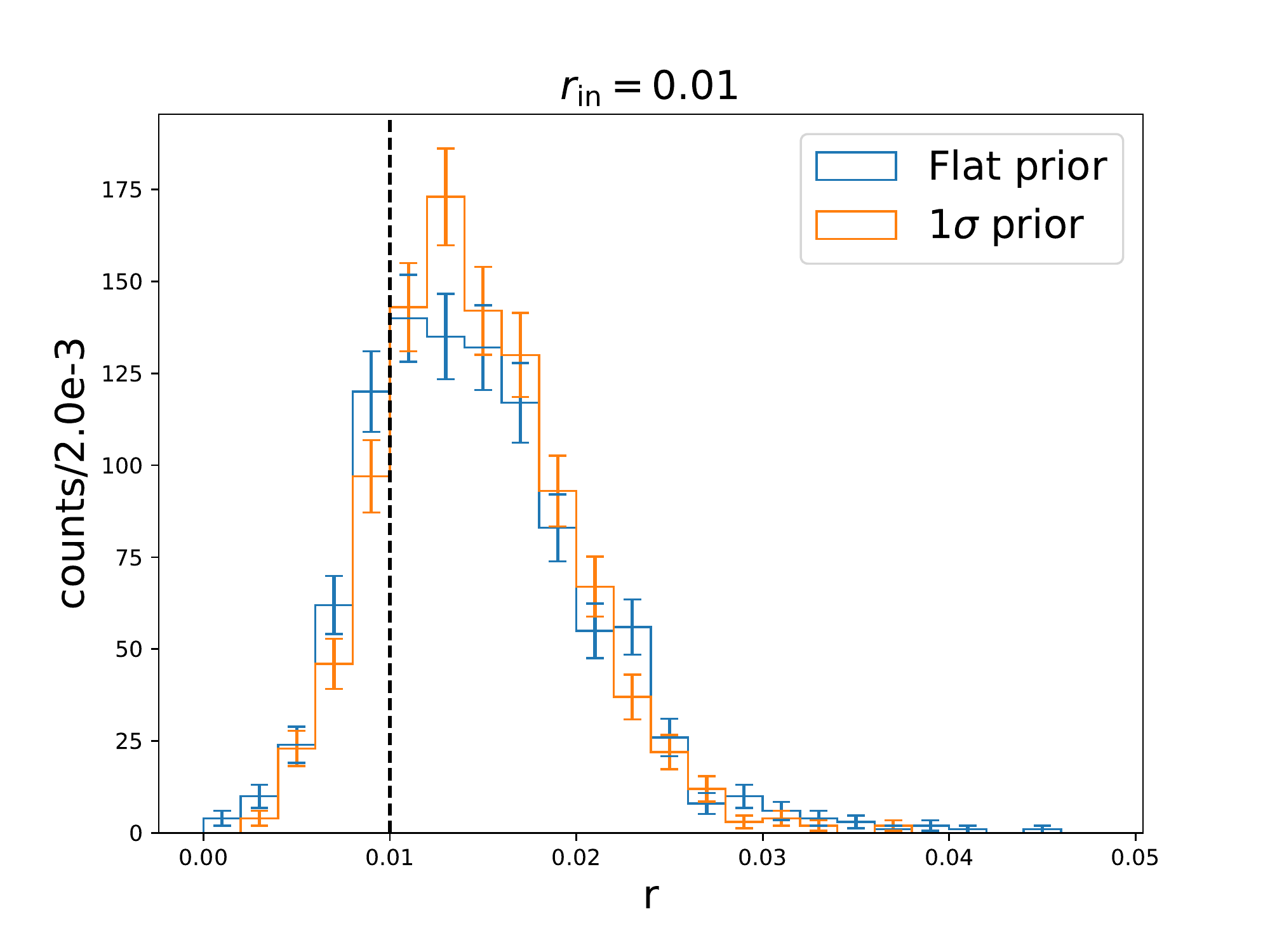}
	\caption{ 
	Histograms of $r_\text{out}$ from $1000$ realizations for a flat prior and a $1\sigma$ prior on $T_d$.
	The input $r_\text{in}=0.01$ and dust model is 1MBB.
	The modified high-frequency model is used.
    \label{fig:Dust_Synch_var_LTD_high_15freq_r1e-2_2plot}}
\end{figure}

We show histograms of $r_\text{out}$ against $r_\text{in}=0.01$ with flat prior and a $1\sigma$ prior in Fig.~\ref{fig:Dust_Synch_var_LTD_high_15freq_r1e-2_2plot} for the modified high-frequency model.
Compared to the result with the fiducial frequency band setting shown in the left panel of Fig.~\ref{fig:Dust_Synch_var_LTD17_15freq}, the uncertainties on $r$ become much smaller while the $50$th percentile values of $r$ are positively biased as shown in Table~\ref{tab:results} if the modified high-frequency model is considered.
This tendency was also found in Ref.~\cite{Ichiki:2018glx};
the smaller uncertainty comes with larger systematic bias if one sets the foreground frequency bands far away from the CMB bands.

\subsection{Mismodeling}
\begin{figure}
	\centering
	\includegraphics[width= 0.48\columnwidth]{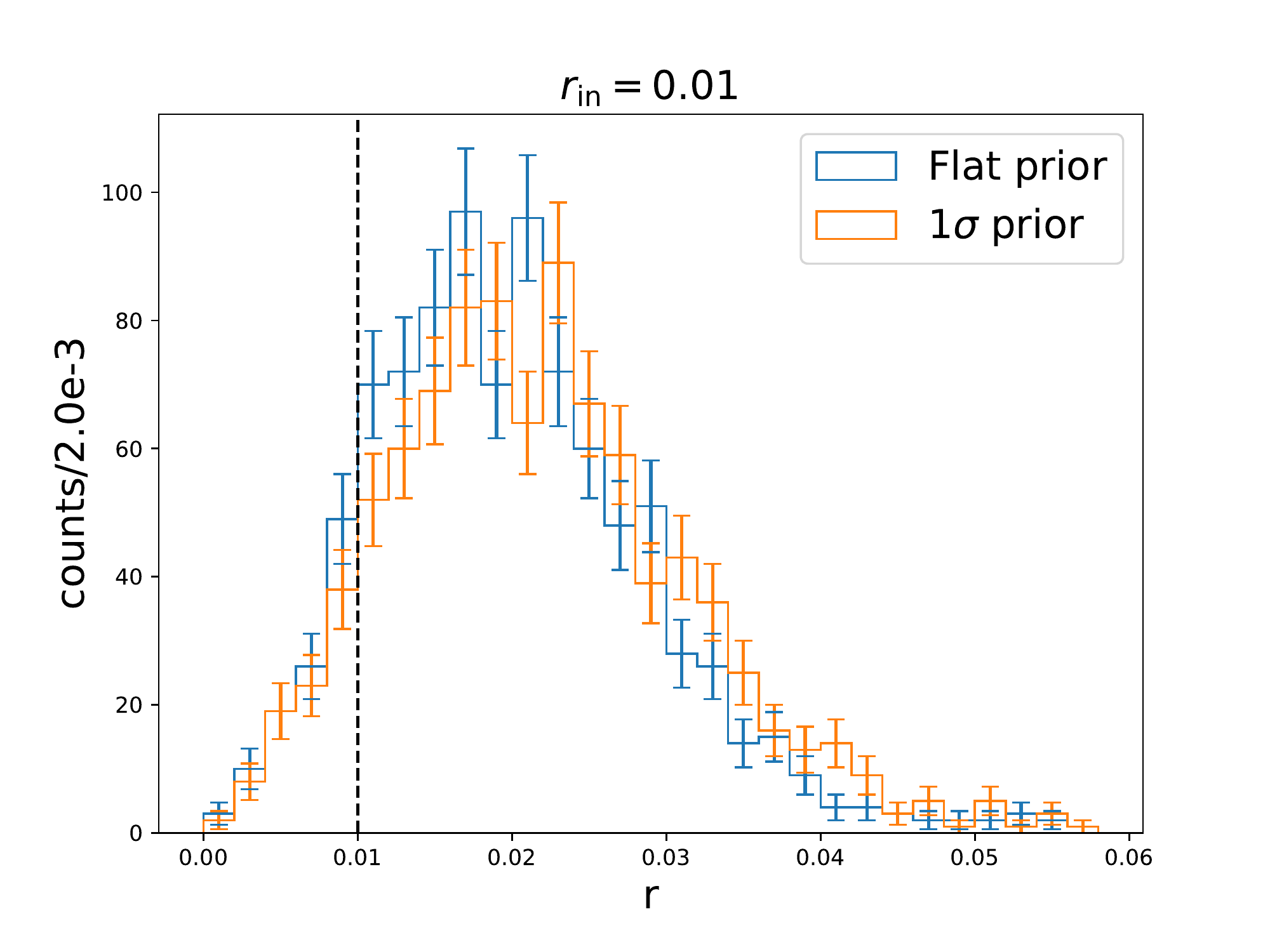}
	\caption{
	Histograms of $r_\text{out}$ from $1000$ realizations.
		The input $r_\text{in}=0.01$ and dust model is 2MBB. \label{fig:Dust_Synch_var_LTD17_15freq_2MBB_r1e-2}}
\end{figure}

To demonstrate the case in which we assume a wrong foreground model,
we use the two-component modified black body (2MBB) model as the input but the 1MBB model to estimate $r$ and the foreground parameters.
We show the histograms of the estimated $r$ from $1000$ realizations with flat and $1\sigma$ priors on $T_d$ in Fig.~\ref{fig:Dust_Synch_var_LTD17_15freq_2MBB_r1e-2}.
We find that the estimates of $r$ are biased by $\sim 0.01$.

This bias was not found in our previous paper~\cite{Ichiki:2018glx}.
The reason could be because the determinative power of the foreground parameters depends on the number of frequency bands.
In Ref.~\cite{Ichiki:2018glx}, $r$ and 1MBB foreground parameters were estimated using only six bands.
Because of the small number of bands used in the analysis, the foreground parameters were not well determined and therefore the bias on $r$ was small.
To confirm this argument, we estimated $r$ with seven and $10$ bands using the extended delta-map method.
The results showed that the bias was small for the case with seven bands, and large with $10$ bands as well as with $15$ bands.

\begin{figure}
	\centering
	\begin{minipage}{0.45\textwidth}
		\centering
		\includegraphics[width=\columnwidth]{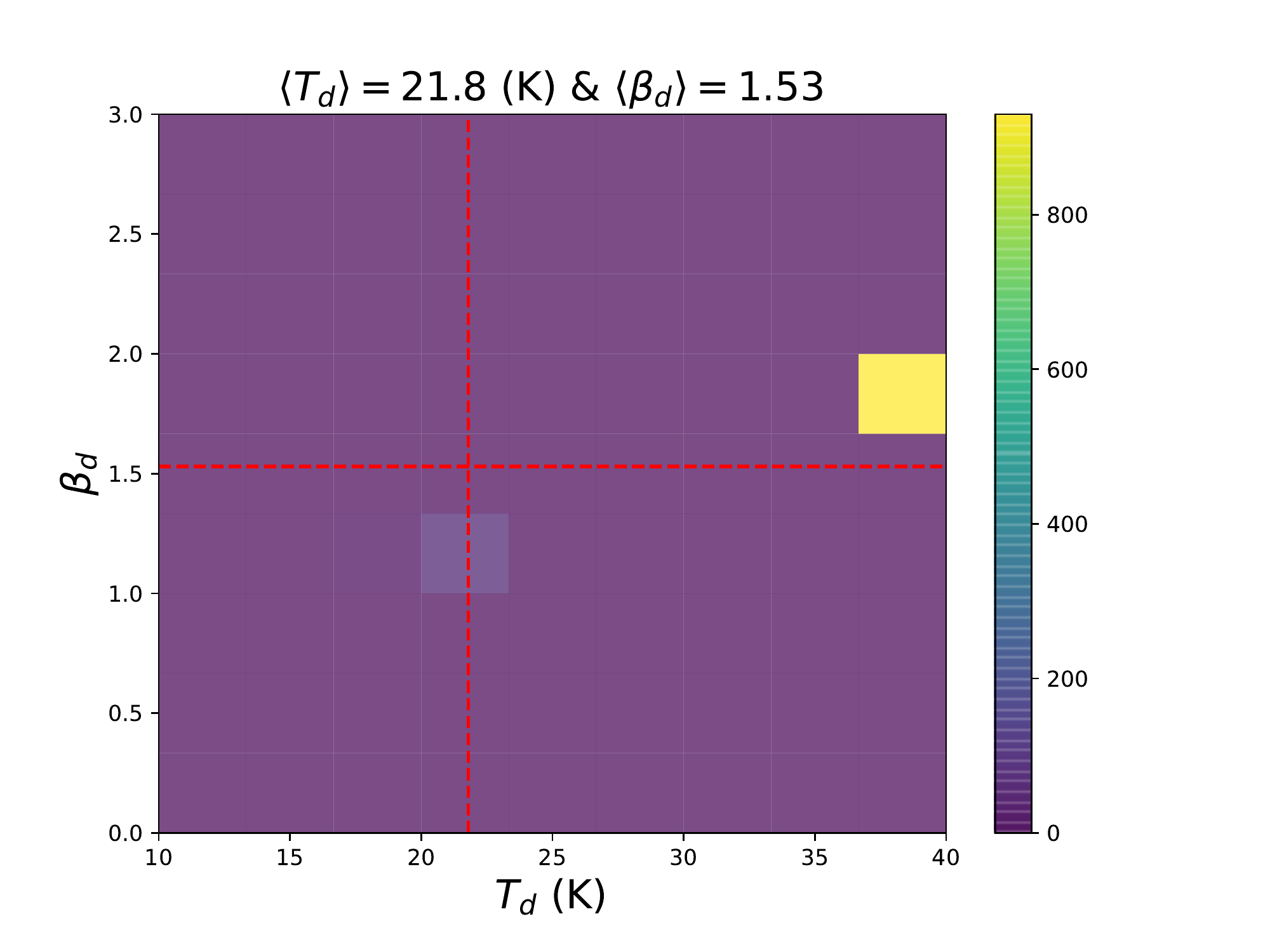}
	\end{minipage}
    \hfill
	\begin{minipage}{0.45\textwidth}
	\centering
	\includegraphics[width=\columnwidth]{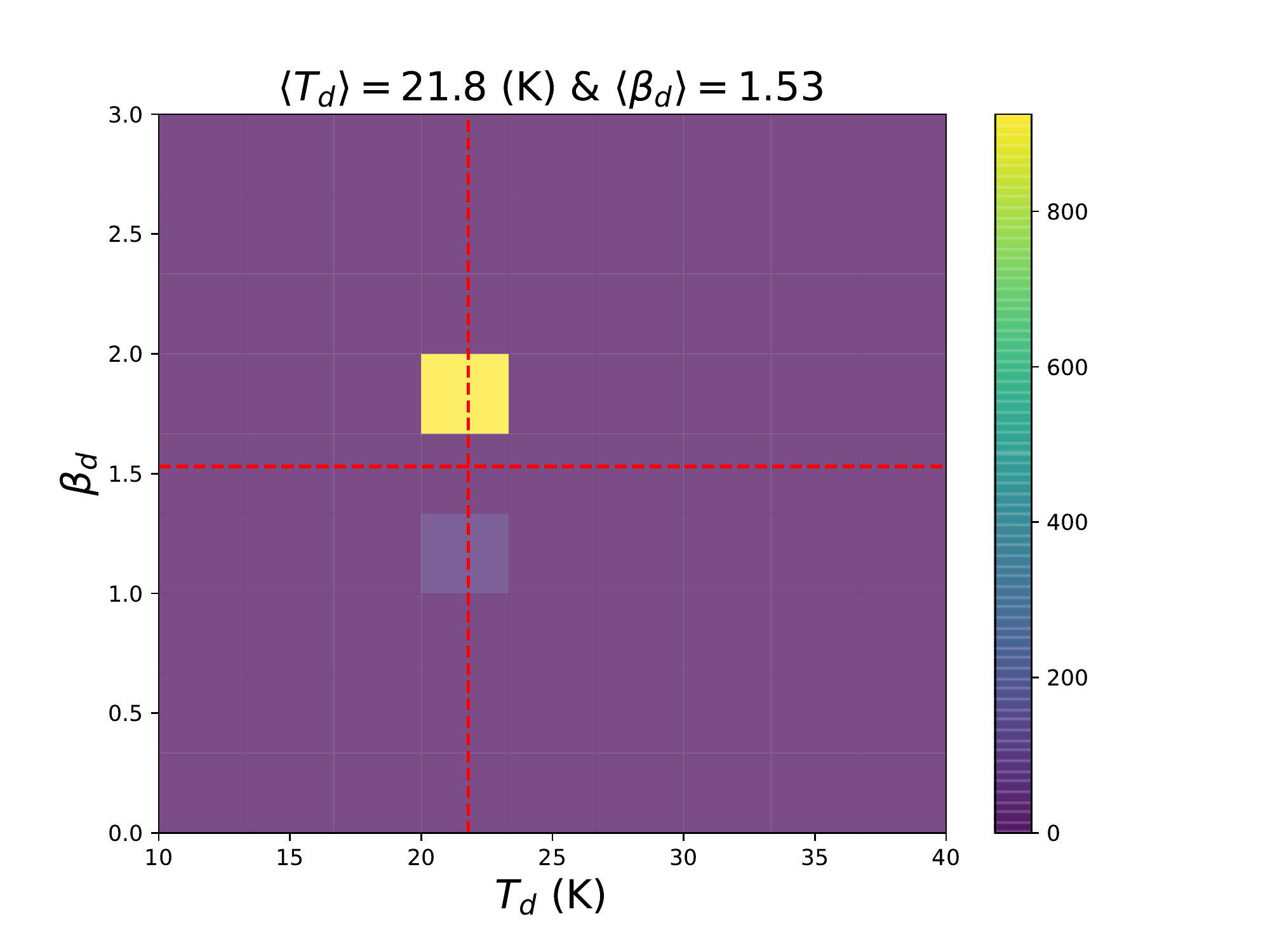}
	\end{minipage}
\caption{\label{fig:Dust_Synch_var_LTD17_15freq_2MBB_r1e-2_Tdpriors_Td_vs_betad}
2D histograms of estimated $T_d$ and $\beta_d$ in the case of mismodeling with flat (left) and $1\sigma$ (right) priors on $T_d$.
}
\end{figure}

We show 2D histograms of foreground parameters, $T_d$ and $\beta_d$, in the left and right panels of Fig.~\ref{fig:Dust_Synch_var_LTD17_15freq_2MBB_r1e-2_Tdpriors_Td_vs_betad} with flat and $1\sigma$ priors on $T_d$, respectively.
Without a prior, estimated $T_d$ reaches the bound.

\section{Summary and discussion\label{sec:conclusion}}
In this paper, we have improved the ``delta-map method''~\cite{Ichiki:2018glx} by constructing a parametric likelihood in a Bayesian way so that more observation bands can be used.
By incorporating the covariance of the foreground emission as ``vague'', 
we have extended the method.
Sample codes are available at  \texttt{extended-deltamap} GitHub repository~\url{https://github.com/YutoMinami/extended-deltamap}.

We have tested the ``extended delta-map method'' with realistic simulations assuming LiteBIRD-like telescopes. 
In the case with one foreground component, we find that the extended delta-map method can estimate both the CMB parameter (tensor-to-scalar ratio $r$) and the foreground parameters, even if we use more frequency bands than the minimum required bands (Sect.~\ref{sec:results_synch} and Sect.~\ref{sec:results_dust}).
Moreover, we also found that the parameters of the foreground model are better determined
when more observation frequency bands are used,
and the tensor-to-scalar ratio $r$ is better estimated accordingly.
This is an improvement and benefit compared to the previous delta-map method.

Next, we apply our method to the model with two foreground components, synchrotron and dust.
It turns out that determining dust foreground parameters $\beta_d$ and $T_d$ becomes difficult in this case.
The reason for this probably lies in the interplay between the synchrotron and dust foreground emissions.
To determine $\beta_d$ and $T_d$ simultaneously,
one needs to observe the dust spectrum over a wide frequency range.
However, in the two-component foreground model,
synchrotron radiation dominates the low-frequency side and masks the dust component, effectively reducing the number of observed frequency bands that can be used to estimate dust foreground parameters.
To aid the determination of the foreground parameter, therefore,
we have imposed a prior on dust temperature based on the Planck results.
In this case, the error in the estimate of $r$ becomes small,
while we have a small positive bias in the estimate.
The small bias may indicate that the mean value of $T_d$ estimated in the delta-map method is not necessarily the same as $T_d$ derived by the maps with higher angular resolutions, e.g., Planck. 

A positive bias is also found in the case using a modified band configuration with higher frequencies.
If we increase the frequencies of the two highest observation bands, 
we can determine the foreground parameters better, as we discussed above, while the estimated CMB parameter is biased.
This tendency was already found in in Ref.~\cite{Ichiki:2018glx}.
The reason for this is probably the breakdown of the perturbative treatment of the foreground parameters in our method.
Although it is easier to remove foreground radiation when the frequency bands are closer to each other,
internal template methods, including ours, have the disadvantage of removing the CMB at the same time, resulting in relatively larger noise.
Thus, we are faced with the familiar dilemma of systematic and statistical errors that are inherent in statistical analysis.

Lastly, we test the case with mismodeling by fitting the two-component modified black body model with the one-component modified black body model.
We find that the estimated CMB parameter is biased.

In this paper, we have not considered detailed characteristics of telescopes or detectors, e.g., bandpass average discussed in Ref.~\cite{Ichiki:2018glx}.
We can include the detailed characteristics in the transfer matrices.
The basic idea of the delta-map method is to consider the spatial variation of foreground signal parameters perturbatively.
Though we have only considered the first-order expansions of spatial variations of foreground parameters,
we can consider higher-order expansions,
which may improve the estimates of the foreground and CMB parameters.
We leave this study for future works.

\section*{Acknowledgment}
We thank E.~Komatsu, A.~Nishizawa, Y.~Chinone, and S.~Takakura for useful discussions.
This work is supported in part by the Japan Society for the Promotion of Science (JSPS) KAKENHI, Grants Nos.~JP18K03616 (K.I.) and JP20K14497 (Y.M.), JSPS core-to-core program number JPJSCCA20200003 (K.I.),
JST AIP Acceleration Research Grant JP20317829 (K.I.),
and JST FOREST Program JPMJFR20352935 (K.I.).

\bibliographystyle{ptephy}
\bibliography{references}
%

\appendix
\section{Derivation of likelihood}\label{sec:derivation}

We use the methodology the methodology described in Ref.~\cite{Rasmussen:2006} to deal with ``vague'' foreground covariance matrix.

When we marginalize $P(\sfg,  \Sfg )$ in \eqref{eq:SNLike} ,  we have
\begin{align}\begin{split}
-2\ln P(\bar{p}^{I}, \sfg, \bm{S}|\vec{m} ) &= 
 \left( \vec{m} - \tilde{\bm{D}}\vec{s_{F}}  \right)^\T	\left( ( \Scmb + \bm{N} )  + \Dfg \Sfg  \Dfg^\T  \right)^{-1}	\left( \vec{m}  -  \Dfg \sfg   \right)
 \\& +
 n \ln 2\pi  + \ln\left| ( \Scmb +\bm{N} ) + \Dfg  \Sfg   \Dfg^\T \right|.
\end{split}\end{align}

When we assume that the mean of foreground signal is zero, $\sfg = \vec{0}$, we have
\begin{align}\begin{split}
-2\ln P(\bar{p}^{I},  \bm{S} | \vec{m} ) &= 
\vec{m} ^\T  ( \Scmb + \bm{N} )^{-1} \vec{m}
\\
&- \vec{m} ^\T 
(\Scmb + \bm{N} )^{-1} \Dfg  \left[  {\Sfg}^{-1} + \Dfg ^\T  ( \Scmb + \bm{N} )^{-1} \Dfg  \right]^{-1}   \Dfg^\T ( \Scmb + \bm{N} )^{-1} 
\vec{m}
\\
& + \ln |\Scmb + \bm{N} | +  \ln | \Sfg | +\ln \left| \Dfg^\T (\Scmb + \bm{N} )^{-1} \Dfg  \right| + \mathrm{const.} \,,
\end{split}\end{align}
where we used \textit{matrix inversion lemma} \eqref{eq:MIL}, ~\eqref{eq:MILlog}.

When we are ignorant about the covariance of the foreground signals, 
we take the limit where $\Sfg^{-1} \rightarrow \bm{O}$ and have
\begin{align}\begin{split}\label{eq:SNFLike}
-2\ln P(\bar{p}^{I}, \Socmb| \vec{m} ) &=
\vec{m} ^\T  ( \Scmb + \bm{N} )^{-1} \vec{m}
\\ &- \vec{m} ^\T 
(\Scmb + \bm{N} )^{-1} \Dfg  \left[    \Dfg ^\T  ( \Scmb + \bm{N} )^{-1} \Dfg  \right]^{-1} \Dfg^\T ( \Scmb + \bm{N} )^{-1} 
\vec{m}
\\
& + \ln |\Scmb + \bm{N} |  + \ln \left| \Dfg^\T (\Scmb + \bm{N} )^{-1} \Dfg  \right| + \mathrm{const.} \,,
\end{split}\end{align}
where we discard the terms $\ln | \Sfg | $ following Ref.~\cite{Rasmussen:2006}.

This is similar to the likelihood function (Eq.~(61) of Ref.~\cite{Ichiki:2018glx}) except for the additional term, $\ln \left| \Dfg (\Scmb + \bm{N} )^{-1} \Dfg  \right|$.
We will revisit this difference in Appendix~\ref{sec:Comparison} and show that this term is necessary to reproduce the results of Ref.~\cite{Ichiki:2018glx}.

Next, we apply \textit{matrix inversion lemma} \eqref{eq:MIL}, ~\eqref{eq:MILlog} to $(\Scmb + \bm{N})^{-1}$ and $\ln|(\Scmb + \bm{N})|$, to reduce the calculation cost of the large covariance matrix.
Using $ (\Scmb + \bm{N} )$ expressed as $( \Dcmb \Socmb {\Dcmb}^\T +\bm{N}) $, 
we have
\begin{align}
	(\Scmb + \bm{N} )^{-1} & = \bm{N}^{-1} - \bm{N}^{-1} \Dcmb \bm{A}^{-1} {\Dcmb}^\T \bm{N}^{-1}  \label{eq:SpNA} \\ 
	\ln|\Scmb + \bm{N}|&= \ln|\Socmb| + \ln|\bm{N}| + \ln|\bm{A}|, \label{eq:SpNAlog}
\end{align}
where we define
$\bm{A} = ( {\Socmb}^{-1} + \sum_{j =1}^{N+1} \bm{N}_{\nu_j}^{-1} )$.
When we substitute Eqs.~\eqref{eq:SpNA} and \eqref{eq:SpNAlog} into Eq.~\eqref{eq:SNFLike},
we have 
\begin{align}\begin{split}
-2\ln P(\bar{p}^{I}, \Socmb | \vec{m} ) &=
\vec{m}^\T  \bm{N}^{-1} \vec{m} 
- \vec{m}^\T \bm{N}^{-1} \Dcmb \bm{A}^{-1} {\Dcmb}^\T \bm{N}^{-1}\vec{m}
		-\vec{M}^\T \bm{H}\vec{M} - \vec{M}^\T \bm{H}\bm{B}^{-1}\bm{H}\vec{M}
		\\& + \ln |\Socmb| + \ln|\bm{N} |+ \ln| { \Dfg }^{T} \bm{N}^{-1} \Dfg  |+ \ln|\bm{B}|,
\end{split}\end{align}
where we define
\begin{align}
	\vec{M} &= \vec{m} - \Dcmb \bm{A}^{-1}  {\Dcmb}^\T \bm{N}^{-1} \vec{m} \label{eq:meanvec}, \\
	\bm{H} &=  \bm{N}^{-1} \Dfg \left[  {\Dfg}^\T \bm{N}^{-1} \Dfg\right]^{-1} {\Dfg}^\T \bm{N}^{-1}, \\
	\bm{B} &= \bm{A} - { \Dcmb}^\T \bm{H}\Dcmb.
\end{align}
Here we try to keep symmetry to write down each term 
because we find that the estimated parameters are biased when we calculate the log-likelihood value using asymmetric terms.
Finally, we summarize newly defined matrices and the vector in Table~\ref{tab:definition}.

\begin{table}
	\centering
\begin{tabular}{c|c}
	\toprule
	Symbols &  Definition\\
	\midrule
	$\bm{A}$& $( {\Socmb}^{-1}   + \sum_{j =1}^{N+1} \bm{N}_{\nu_j}^{-1} )$\\
	$\bm{H}$& $\bm{N}^{-1} \Dfg \left[  {\Dfg}^\T \bm{N}^{-1} \Dfg\right]^{-1} {\Dfg}^\T\bm{N}^{-1}$\\
	$\bm{B}$&$\bm{A} - { \Dcmb}^\T \bm{H}\Dcmb$\\
	$\vec{M}$&$\vec{m} - \Dcmb \bm{A}^{-1}  {\Dcmb}^\T \bm{N}^{-1} \vec{m}$\\
	\bottomrule
\end{tabular}
\caption{\label{tab:definition} 
Summaries of matrices and the vector defined in this section. 
}
\end{table}

\section{\label{sec:lemma}Matrix inversion lemma}
In the derivation of the likelihood function, we use Woodbury, Sherman, and  Morrison formula, so-called ``matrix inversion lemma'',
\begin{align} \label{eq:MIL}
\left( Z + UWV^\T \right)^{-1} = Z^{-1} - Z^{-1}U\left( W^{-1} + V^\T Z^{-1} U \right)^{-1}VZ^{-1},
\end{align}
where $Z$ and $W$ are the invertible matrices and $U$ and $V$ are matrices with corresponding dimensions.

For log-determinants, similar equation exists:
\begin{align}\label{eq:MILlog}
\ln \left|  Z + UWV^\T \right| = \ln|Z| + \ln |W| +\ln| W^{-1} + V^\T Z^{-1} U | .
\end{align}
\section{Comparison of likelihood}\label{sec:Comparison}
In this section, 
we show that the estimated parameters with Eq.~\eqref{eq:SNLikeFinal} are equivalent to those estimated from Eq.~\eqref{eq:IchikiLike} in the case in which we assume one foreground component and one parameter, $\beta$, where the number of frequencies is $N_\nu= \left( N_d(=1) + 1\right) + 1=3$.

The cleaned CMB map~\eqref{eq:cleanedCMBalpha} can also be expressed as 
$\left[ \bm{D}^{-1}\vec{m}\right]_\mathrm{CMB}$, where the subscript ``CMB'' means that we take the elements related to the CMB, as described above Eq.~(50) in Ref.~\cite{Ichiki:2018glx}.

First, let us calculate this cleaned CMB map.
The elements of $\bm{D}$ matrix are as 
\begin{align}\begin{split}
	\bm{D}  &= \begin{bmatrix}\Dcmb & \Dfg \end{bmatrix}  \\
	&= \begin{bmatrix}
		\bm{I} & \textsl{g}_{\nu_1} D_{\nu_1} \bm{I}& \textsl{g}_{\nu_1} D_{\nu_1,\beta}\bm{I} \\
		\bm{I} & \textsl{g}_{\nu_2} D_{\nu_2} \bm{I}& \textsl{g}_{\nu_2} D_{\nu_2,\beta} \bm{I}\\
		\bm{I} & \textsl{g}_{\nu_3} D_{\nu_3} \bm{I}& \textsl{g}_{\nu_3} D_{\nu_3,\beta} \bm{I}
	\end{bmatrix}.
\end{split}\end{align}
The inverse matrix of $\bm{D}$ is as
\begin{align}\begin{split}
	\bm{D}^{-1}  
	&=\frac{-1}{  |\bm{\Xi}    \Dcmb| } \begin{bmatrix}
		-\xi_{23} \bm{I}&  - \xi_{31} \bm{I}&  -\xi_{12}  \bm{I} \\
        (\textsl{g}_{\nu_2} D_{\nu_2,\beta} - \textsl{g}_{\nu_3} D_{\nu_3})   \bm{I}&  ( -\textsl{g}_{\nu_1} D_{\nu_1,\beta}  + \textsl{g}_{\nu_3} D_{\nu_3,\beta} )  \bm{I} & (\textsl{g}_{\nu_1} D_{\nu_1,\beta} - \textsl{g}_{\nu_2} D_{\nu_2,\beta} ) \bm{I} \\
        ( - \textsl{g}_{\nu_2} D_{\nu_2}  + \textsl{g}_{\nu_3} D_{\nu_3} ) \bm{I} & ( \textsl{g}_{\nu_1} D_{\nu_1} - \textsl{g}_{\nu_3} D_{\nu_3} )\bm{I}&  (- \textsl{g}_{\nu_1} D_{\nu_1}  +  \textsl{g}_{\nu_2} D_{\nu_2} )\bm{I}
	\end{bmatrix},
\end{split}\end{align}
where $\xi_{ij} = \textsl{g}_{\nu_i}\textsl{g}_{\nu_j} (D_{\nu_i,\beta} D_{\nu_j} - D_{\nu_i}D_{\nu_j,\beta} )$ and  $\bm{\Xi} = \begin{pmatrix}
	\xi_{23}\bm{I}&\xi_{31}\bm{I}&\xi_{12}\bm{I}
\end{pmatrix} $ .
Then the cleaned CMB map can be expressed as
\begin{align}\begin{split}
	\mathrm{CMB}^\mathrm{ML}(\hat{n}) &= \left[ \bm{D}^{-1}\vec{m}\right]_\mathrm{CMB}\\
	&= (\bm{\Xi}    \Dcmb)^{-1} \bm{\Xi} \vec{m} .
	\end{split}
\end{align}
The corresponding covariance matrix~\eqref{eq:cov_Ichiki} is calculated as
\begin{align}\label{eq:IchikiLike}
	\bm{C} =  \Socmb + |\bm{K}|(\bm{\Xi}    \Dcmb)^{-1}(\bm{\Xi}    \Dcmb)^{-1},
\end{align}
where $\bm{K} = \bm{\Xi} \bm{N} \bm{\Xi}^\T $.

Therefore, we can write the likelihood function used in Ref.~\cite{Ichiki:2018glx} as
\begin{align}\begin{split}
		-2\ln\mathcal{L}  =   ((\bm{\Xi}    \Dcmb)^{-1} \bm{\Xi} \vec{m} )^\T \left( \Socmb  + |\bm{K}| (\bm{\Xi}    \Dcmb)^{-1}(\bm{\Xi}    \Dcmb)^{-1} \right)^{-1}  (\bm{\Xi}    \Dcmb)^{-1} \bm{\Xi} \vec{m}  .
\end{split}\end{align}
We will show that Eq.~\eqref{eq:SNLikeFinal} is equal to this equation.

First, we calculate ${\Dfg}^\T  \bm{N}^{-1}  \Dfg $ matrix and its determinant as 
\begin{align}
	{\Dfg}^\T  \bm{N}^{-1}  \Dfg  &=
	\begin{bmatrix}
		\sum_i  N_i^{-1} \textsl{g}_{\nu_i}^2 D_{\nu_i}^2  &  \sum_i N_i^{-1} \textsl{g}_{\nu_i} D_{\nu_i}D_{\nu_i,\beta}  \\
		 \sum_i N_i^{-1} \textsl{g}_{\nu_i} D_{\nu_i}D_{\nu_i,\beta}  & \sum_i  N_i^{-1} \textsl{g}_{\nu_i}^2 D_{\nu_i,\beta}^2 
	\end{bmatrix}
	\\
	\left| \tilde{\bm{D}}^{\rm T}\bm{N}^{-1}\tilde{\bm{D}} \right|&=  |\bm{K}| \prod_{i} |N_i^{-1}| . 
\end{align}
Using them,  we have 
\begin{align}\begin{split}
		\bm{H} &=\frac{1}{|\bm{K}| } \begin{bmatrix}
			N_1^{-1 }(N_3 \xi_{12}^2 +  N_2 \xi_{31}^2 )\bm{I} & -\xi_{31}\xi_{23} \bm{I} &-\xi_{12}\xi_{23}\bm{I} \\
			-\xi_{31}\xi_{23}\bm{I} & N_1^{-2 }(N_1 \xi_{23}^2 +  N_3 \xi_{12}^2 )\bm{I} & -\xi_{12}\xi_{31}\bm{I} \\
			-\xi_{12}\xi_{23}\bm{I} & -\xi_{12}\xi_{31}\bm{I} &N_1^{-3 }(N_2 \xi_{31}^2 +  N_1 \xi_{23}^2 )\bm{I} 
		\end{bmatrix}\\
		&=\bm{N}^{-1} - \frac{ \bm{\Xi}^\T  \bm{\Xi} }{|\bm{K}|}.
\end{split}\end{align}
Then we have
\begin{align}\begin{split}
		\bm{B} &= \bm{A} -   {\Dcmb}^\T  \left(\bm{N}^{-1} - \frac{ \bm{\Xi}^\T  \bm{\Xi} }{|\bm{K}|}\right) \Dcmb  \\
		&= { \Socmb}^{-1}  +  \frac{{\Dcmb}^\T  \bm{\Xi}^\T  \bm{\Xi} \Dcmb}{|\bm{K}|},
\end{split}\end{align}
and its inverse as 
\begin{align}\begin{split}
		\bm{B}^{-1} &= \left({ \Socmb}^{-1}  +  \frac{{\Dcmb}^\T  \bm{\Xi}^\T  \bm{\Xi} \Dcmb}{|\bm{K}|}\right)^{-1}\\
		&= |\bm{K}|(\bm{\Xi} \Dcmb)^{-1} (\bm{\Xi} \Dcmb)^{-1} \Socmb \left( { \Socmb}  +  |\bm{K}|(\bm{\Xi} \Dcmb)^{-1} (\bm{\Xi} \Dcmb)^{-1} \right)^{-1}
		\\&=
		 |\bm{K}|(\bm{\Xi} \Dcmb)^{-1} (\bm{\Xi} \Dcmb)^{-1} \Socmb \bm{C}^{-1}
		 \\(&= \bm{A}^{-1} +\bm{B}^{-1} {\Dcmb}^\T \bm{H}\Dcmb \bm{A}^{-1})
		 \\(&= \bm{A}^{-1} +\bm{A}^{-1} {\Dcmb}^\T \bm{H}\Dcmb \bm{B}^{-1}).
\end{split}\end{align}

Therefore, we can rewrite Eq.~\eqref{eq:SNLikeFinal} as
\begin{align}\begin{split}
		-2\ln P(\bar{p}^{I}, \Socmb | \vec{m} ) &=
		\vec{m}^\T  \bm{N}^{-1} \vec{m} 
		- \vec{m}^\T \bm{N}^{-1} \Dcmb \bm{A}^{-1} {\Dcmb}^\T \bm{N}^{-1}\vec{m}
		- \vec{M}^\T \bm{H}\vec{M} 
		\\&- \vec{M}\bm{H}\Dcmb \bm{B}^{-1} {\Dcmb}^\T \bm{H}\vec{M}
		\\& + \ln |\Socmb| + \ln|\bm{N} |+ \ln| { \Dfg }^{T} \bm{N}^{-1} \Dfg  |+ \ln|\bm{B}|
		\\
		&= 	\vec{m}^\T  \bm{N}^{-1} \vec{m} 
		- \vec{m}^\T \bm{N}^{-1} \Dcmb \bm{A}^{-1} {\Dcmb}^\T \bm{N}^{-1}\vec{m}
		- \vec{m^\T }\bm{H}\vec{m} 
		\\&
		+2 \vec{m^\T }\bm{H}\Dcmb \bm{A}^{-1} {\Dcmb}^\T \bm{N}^{-1} \vec{m} 
		\\&
		- \vec{m}^\T \bm{N}^{-1} \Dcmb \bm{A}^{-1} {\Dcmb}^\T \bm{H}\Dcmb \bm{A}^{-1} {\Dcmb}^\T \bm{N}^{-1} \vec{m} 
		\\&
		- \vec{m}^\T \bm{H}\Dcmb\bm{B}^{-1}{\Dcmb}^\T \bm{H}\vec{m}
		+2 \vec{m}\bm{H}{\Dcmb}\bm{B}^{-1}{\Dcmb}^\T \bm{H} {\Dcmb} \bm{A}^{-1} {\Dcmb}^\T \bm{N}^{-1}\vec{m} 
		\\&-  \vec{m}\bm{N}^{-1} {\Dcmb} \bm{A}^{-1} {\Dcmb}^\T \bm{H}{\Dcmb} \bm{B}^{-1}{\Dcmb}^\T \bm{H} {\Dcmb}\bm{A}^{-1} {\Dcmb}^\T \bm{N}^{-1}\vec{m}
				\\& + 2\ln|\bm{\Xi} \Dcmb | +\ln|\bm{C}|
	\\
	&= \vec{m}^\T (\bm{\Xi} \Dcmb)^{-1}\frac{\bm{I} - \Socmb \bm{C}^{-1}}{|\bm{K}|}(\bm{\Xi} \Dcmb)^{-1}\vec{m}  + 2\ln|\bm{\Xi} \Dcmb | +\ln|\bm{C}|
	\\&= \mathrm{CMB}^\mathrm{ML}(\hat{n})^\T \bm{C}^{-1} \mathrm{CMB}^\mathrm{ML}(\hat{n}) + 2\ln|\bm{\Xi} \Dcmb | +\ln|\bm{C}|.
\end{split}\end{align}
This agrees to Eq.~\eqref{eq:IchikiLike} except for the term, $\ln|\bm{\Xi} \Dcmb |$.
This does not change the estimate parameters,
since, in Ref.~\cite{Ichiki:2018glx}, foreground parameters are determined only with chi-squared term and CMB parameters are determined with the total likelihood function by fixing foreground parameter.

\end{document}